\documentclass[11pt]{article} 
\usepackage{setspace}

\usepackage[utf8]{inputenc} 
\usepackage{amsmath}
\usepackage{amssymb}
\usepackage[colorlinks,
linkcolor=blue,
anchorcolor=blue,
citecolor=blue,
]{hyperref}

\usepackage{geometry} 
\usepackage{graphicx} 

\usepackage{booktabs} 
\usepackage{array} 
\usepackage{paralist} 
\usepackage{verbatim} 
\usepackage{subfig} 
\usepackage{amsmath} 
\usepackage{fancyhdr} 
 \usepackage[numbers,sort&compress]{natbib} 
\usepackage{sectsty}
\allsectionsfont{\sffamily\mdseries\upshape} 
\usepackage{indentfirst}
\usepackage[nottoc,notlof,notlot]{tocbibind} 
\usepackage[titles,subfigure]{tocloft} 
\usepackage{authblk}
\usepackage{abstract}
 
\title{Thermal Metamaterials for Temperature Maintenance: From Advances in Heat Conduction to Future Convection Prospects}
\author{Xinchen Zhou \thanks{20110190014@fudan.edu.cn.}}
\affil{Department of Physics, State Key Laboratory of Surface Physics, Key Laboratory of Micro and Nano Photonic Structures (MOE), Fudan University, Shanghai 200438, China}
\begin{document}
\maketitle

\begin{abstract}
Maintaining temperature is crucial in both daily life and industrial settings, ensuring human comfort and device functionality. In the quest for energy conservation and emission reduction, several contemporary passive temperature control technologies have emerged, including phase change temperature control, shape memory alloys, solar thermal utilization, sky radiation cooling, and heat pipe systems. However, there is a pressing need for more quantitative methods to further optimize temperature maintenance. With advancements in theoretical thermotics and the emergence of thermal metamaterials, it is clear that temperature fields can be precisely manipulated by fine-tuning thermal and structural parameters. This review introduces three innovative devices: the energy-free thermostat, the negative-energy thermostat, and the multi-temperature maintenance container. All are grounded in the principles of thermal metamaterials and primarily operate under conduction heat transfer conditions. When compared with traditional technologies, the unparalleled efficacy of thermal metamaterials in temperature management is evident. Moreover, brief prospects present strategies to improve temperature maintenance under convection heat transfer, thus expanding the application spectrum of thermal metamaterials. This review concludes by spotlighting the evolution and interplay of the aforementioned three devices, marking the progression of thermal metamaterials from theoretical ideas to tangible engineering solutions. These insights not only bridge the gap between applied physics and engineering but also underscore the practical potential of thermal metamaterials.
\end{abstract}
\clearpage
\setstretch{1.1}
\section{Introduction}

Temperature maintenance plays a pivotal role in our daily lives and industrial applications. In building environments, it ensures human comfort, while in battery thermal management, it underpins the performance and safety of the battery. Given the increasing emphasis on energy-saving and emission reduction, achieving efficient temperature maintenance sustainably has never been more paramount \cite{XCZhou-GaoYAE23, XCZhou-ZhangQNC22, XCZhou-WangXRE22}.

As a main part of temperature control, temperature maintenance processes can be categorized into two main types based on their energy requirements: active and passive. Notably, passive temperature maintenance has emerged as a significant player in energy-saving temperature control strategies. Cutting-edge technologies such as phase change temperature control \cite{XCZhou-ChenZJES22,XCZhou-GuoRMSE22}, shape memory alloy \cite{XCZhou-WangBATE23,XCZhou-ReganyDATE22}, solar thermal utilization \cite{XCZhou-LiuWJES22,XCZhou-PaulJ22}, sky radiation cooling \cite{XCZhou-LiuJCEJ23, XCZhou-LiuJAFM22}, and heat pipe systems \cite{XCZhou-ParkCIJHMT22,XCZhou-LengZE22}, have showcased exemplary performance in efficiently regulating heat flow between objects and their surroundings. The underlying principle of these technologies hinges on the self-regulation capability of their thermal/structural properties.

\begin{figure}[h!]
\centering
\includegraphics[width=\textwidth]  {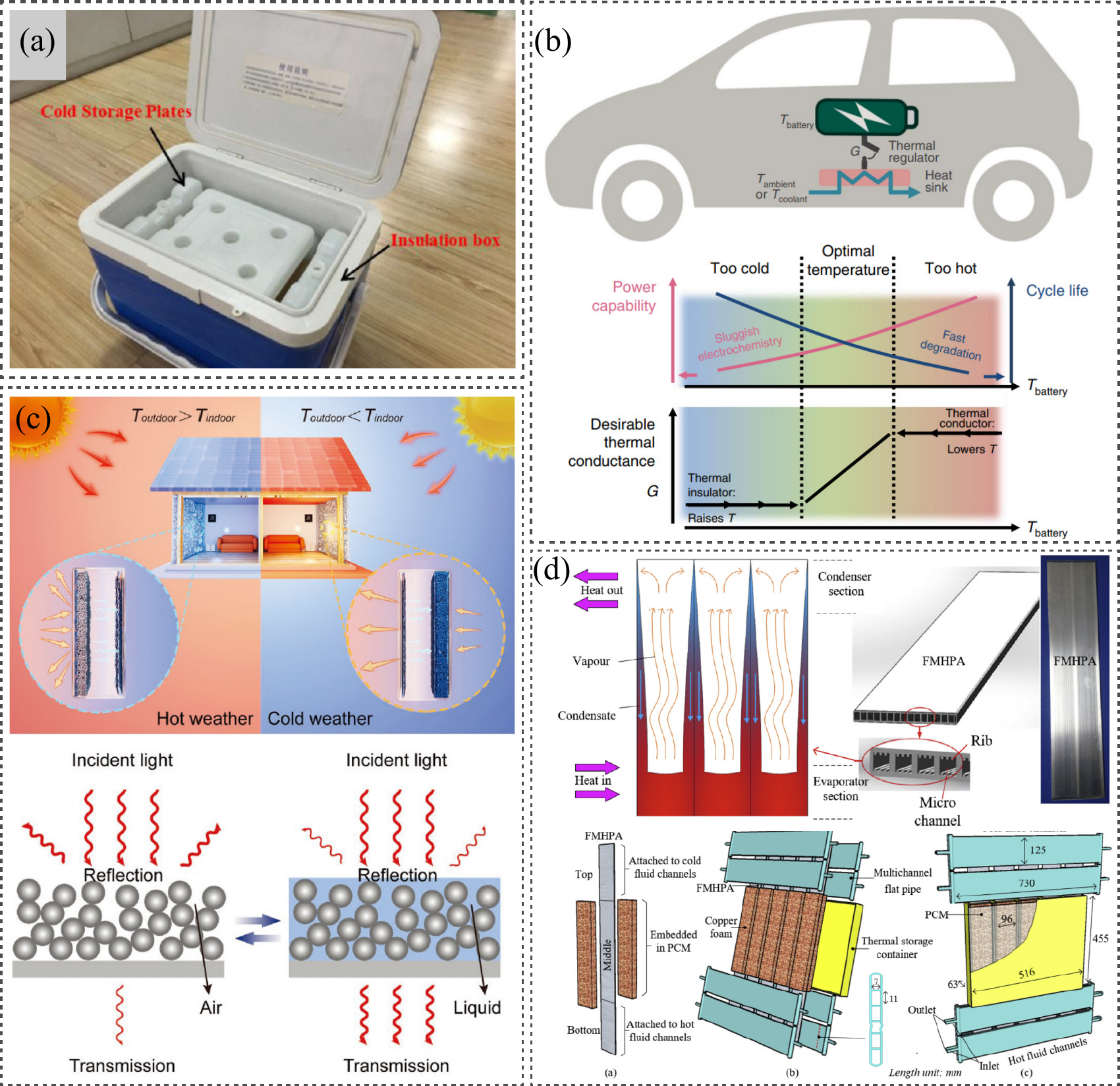}
\caption{\label{Fig. ZXC1}Various technologies for passive temperature maintenance. \textbf{a} Phase change-based temperature control in cold chain logistics. \textbf{b} Use of shape memory alloys for battery thermal management. \textbf{c} Solar thermal utilization combined with sky radiative cooling for thermal comfort. \textbf{d} Integration of phase change and heat pipe technology for waste heat recovery. Adapted from Refs. \cite{XCZhou-DuJJES20,XCZhou-HaoMNE18,XCZhou-ZhangCAFM22,XCZhou-KhalilmoghadamPRE21}.  \textcircled{c} 2020 Elsevier, 2018 Springer Nature, 2022 Wiley‐VCH GmbH.}
 \end{figure}

In the realm of phase change temperature maintenance, phase change materials (PCMs) are deployed to modulate an object's temperature, leveraging the near-constant temperature maintained during the phase transition process. Consider the realm of cold chain logistics as an illustrative example (Fig. \ref{Fig. ZXC1}a): the temperature of transported goods is regulated by phase change materials during their phase transition, all contained within thermally insulating packages \cite{XCZhou-DuJJES20}.

Shape memory alloys (SMAs) are another fascinating domain. The structural changes they undergo before and after phase transitions can alter the system's effective thermal properties, enabling temperature regulation. A case in point is explored in Ref. \cite{XCZhou-HaoMNE18}. Here, an SMA with an optimal phase change temperature is utilized to regulate the heat exchange between a battery and its environment. In high-temperature conditions, the SMA constricts, enhancing heat flow, whereas in cooler environments, it expands, reducing heat flow.

Both solar thermal utilization and sky radiation cooling strategies involve modulating the radiation heat transfer rate between objects and their surroundings to control object temperature. An innovative approach to achieving thermal comfort in building environments, presented in Ref. \cite{XCZhou-ZhangCAFM22}, utilizes a liquid-vapor phase transition. When exposed to high outdoor temperatures, the working fluid in the thermal management device assumes a gaseous form, diminishing its transmittance and thereby cooling the indoor environment. Conversely, in colder conditions, the fluid transitions to a liquid state, increasing its transmittance and warming the interior.

Heat pipe technology similarly exploits vapor-liquid phase transitions. Ref. \cite{XCZhou-KhalilmoghadamPRE21} unveils a pioneering application that marries phase change technology with heat pipe systems to recuperate industrial waste heat. This intricate setup comprises two main components: a micro heat pipe array and a phase change energy storage device. The micro heat pipe array is designed to channel heat from the hot end (representing industrial waste heat sources) to the cold end (representing user heating requirements). The phase change energy storage device's role is to harness surplus waste heat. When an abundance of waste heat is present, the material melts, storing this thermal energy. However, in the absence or scarcity of waste heat, the material solidifies, liberating the previously stored thermal energy, which is subsequently transferred to the cold end via the micro heat pipe array. This cycle ensures a stable temperature for the heat-carrying fluid channeled to the user end.

It is clear that the applications mentioned utilize the phase transition properties of materials to achieve temperature regulation by qualitatively adjusting the heat transfer rate between the object and its surroundings. From an energy conservation standpoint, strict temperature maintenance is only achievable when the net heat flow across the temperature control zone (i.e., the object) is zero. Developing more quantitative analytical tools to explore passive temperature maintenance technologies could lead to improved temperature control efficiency.

In recent times, the advent of thermal metamaterials has shown promise in the realm of passive temperature control, given their proficiency in modulating temperature fields \cite{XCZhou-WangiScience20,XCZhou-DaiJNU21,XCZhou-ZhangNRP23,XCZhou-XuBook23,XCZhou-YangPR21,XCZhou-HuangPhysics20,XCZhou-HuangESEE20,XCZhou-Huang20,XCZhou-FanAPL2008,XCZhou-YangFBRMP23}. This advancement is rooted in the evolution of theoretical thermotics since 2008 \cite{XCZhou-FanAPL2008}. A range of thermal functionalities have been demonstrated, showcasing the vast potential of this innovative tool in temperature regulation \cite{XCZhou-WangiScience20,XCZhou-HuangESEE20,XCZhou-HuangPhysics20,XCZhou-HuangPP18,XCZhou-JiIJMPB18,XCZhou-Tan20}.

For instance, in a display of temperature field manipulation, a thermal metamaterial named the ``thermal cloak" was introduced \cite{XCZhou-YaoISci22,XCZhou-DaiPRAP22,XCZhou-WangPRAP21,XCZhou-WangATE21,XCZhou-WangEPL21,XCZhou-XuIJHMT21,XCZhou-XuCPL20,XCZhou-XuPRE20,XCZhou-YangJAP20,XCZhou-XuPRAP19,XCZhou-YangESEE19,XCZhou-YangJAP19,XCZhou-DaiJAP18,XCZhou-WangJAP18,XCZhou-DaiPRE18,XCZhou-Tan4,XCZhou-QiuIJHT14,XCZhou-QiuEPL13,XCZhou-GaoJAP2009,XCZhou-XuEPL201,XCZhou-LiJJAP10}. It essentially renders objects invisible within a classical core-shell model \cite{XCZhou-HuangPhysics20,XCZhou-FanAPL2008,XCZhou-LiJJAP10,XCZhou-XuBook23,XCZhou-Huang20}. Heat seamlessly flows from the high-temperature boundary to the low-temperature boundary across the shell. As a result, the internal object remains thermally unaffected, while the external temperature distribution remains undisturbed by the object's presence. Following this breakthrough, a slew of functionalities has been introduced, expanding the scope of thermal metamaterials. Notable examples include the thermal expander \cite{XCZhou-HuangPhysics20,XCZhou-Tan20,XCZhou-HuangPB17}, thermal concentrator \cite{XCZhou-ZhuangSCPMA22, XCZhou-Tan4, XCZhou-Tan11},  thermal rotator \cite{XCZhou-YangPRAP20}, thermal sensor \cite{XCZhou-JinIJHMT21,XCZhou-XuEPL20,XCZhou-JinIJHMT20}, thermal illusion \cite{XCZhou-XuEPJB191,XCZhou-XuEPJB17,XCZhou-QiuAIP Adv.15,XCZhou-WangPRAP20}, thermal transparency \cite{XCZhou-XuPRA19a, XCZhou-LiuJAP21, XCZhou-LiuJAP212}, etc \cite{XCZhou-WangJAP17,XCZhou-Tan15}. The innovation did not stop there, with considerations for nonlinear effects \cite{XCZhou-DaiIJHMT20,XCZhou-DaiEPJB18,XCZhou-SuEPL20, XCZhou-DaiJNU21,XCZhou-YangPRE19,XCZhou-ZhuangIJMSD23,XCZhou-JinPNAS23,XCZhou-QuEPL21,XCZhou-LeiEPL21,XCZhou-ZhuangPRE22}, multi-/dual-function \cite{XCZhou-Tan4,XCZhou-ZhuangIJMSD23,XCZhou-JinPNAS23,XCZhou-QuEPL21,XCZhou-LeiEPL21},  intelligence \cite{XCZhou-ZhangPRA23,XCZhou-XuSCPMA20,XCZhou-YangEPL19,XCZhou-XuPRA19,XCZhou-XuEPL19,XCZhou-XuEPJB19,XCZhou-XuPLA18,XCZhou-XuJAP18,XCZhou-WangIJTS18,XCZhou-YangEPL19-1}, programmability \cite{XCZhou-LeiIJHMT23,XCZhou-ZhouESEE19}, spatiotempratally modulation \cite{XCZhou-YangPRA23,XCZhou-LeiMTP23,XCZhou-XuPRL22-1,XCZhou-YangPRAP22,XCZhou-XuPRL22,XCZhou-XuPRE21}, thermal wave systems \cite{XCZhou-XuIJHMT20}, thermal dipole \cite{XCZhou-XuEPJB20,XCZhou-XuPRE19-01}, and other novel principles/functions \cite{XCZhou-DaiPR23,XCZhou-XuPANS23,XCZhou-XuNSR23,XCZhou-LiPF22,XCZhou-ZhangTSEP21,XCZhou-DaiiScience21,XCZhou-XuPRE18,XCZhou-XuEPJB18,XCZhou-JiCTP18,XCZhou-ShangIJHMT18,XCZhou-YangAPL17,XCZhou-ShenAPL16,XCZhou-QiuEur.Phys.J.Appl.Phys.15}.

Examining the triad of fundamental heat transfer modes - heat conduction, heat convection, and heat radiation - it is found that the evolution of thermal metamaterials began with systems focused on heat conduction. This subsequently expanded to include coupled heat convection systems and eventually integrated systems addressing radiant heat \cite{XCZhou-WangCPB22,XCZhou-XuEPL21-01,XCZhou-XuAPL21,XCZhou-XuCPLEL20,XCZhou-XuAPL20,XCZhou-WangPRE20,WangICHMT20,XCZhou-XuEPL21,XCZhou-XuPRAP20,XCZhou-ShangAPL18}. The upshot is a comprehensive capability to control heat flow across the entire spectrum of heat transfer processes. This relates to so-called overall heat transfer, omnithermotics, and omnithermal metamaterials \cite{XCZhou-XuESEE20,XCZhou-WangPRAP20}. Since the thermal parameters of the system were calculated by theory, this might lead to difficulty in finding the materials in nature. To address this obstacle, the effective medium theory was employed for fabricating thermal metamaterials \cite {XCZhou-ZhouEPL23,XCZhou-LinSCPMA22,XCZhou-HuangAMT22,XCZhou-TianIJHMT21,XCZhou-XuPRE19,XCZhou-ShangJHT18,XCZhou-YangJAP07,XCZhou-HuangPR06a,XCZhou-HuangJOSAB05,XCZhou-GaoPRB04,XCZhou-HuangCTP01-1}. Consequently, contemporary thermal metamaterials, bolstered by robust analytical theories, sophisticated design methodologies, and state-of-the-art experimental apparatuses, exhibit immense potential for real-world applications in passive temperature control.

In the realm of temperature control, the development of temperature maintenance is vital for energy conservation. This review showcases three devices rooted in thermal metamaterials designed for conduction heat transfer: an energy-free thermostat utilizing temperature trapping theory with SMA; a groundbreaking negative-energy thermostat that produces electrical energy by merging thermotics with electricity; and a versatile multi-temperature maintenance container, deploying adaptive controls and PCMs to cater to diverse transportation needs. Building upon these innovations, the review also hints at the future of temperature maintenance in convection heat transfer scenarios. Overall, the insights offered here, viewed through the lens of thermal metamaterials, illuminate pathways for enhancing temperature maintenance methodologies.

\section{Developments in conduction heat transfer system}
\subsection{Energy-free thermostat}
\subsubsection{Principle and function}
\begin{figure}[h!]
\centering
\includegraphics[width=6cm]  {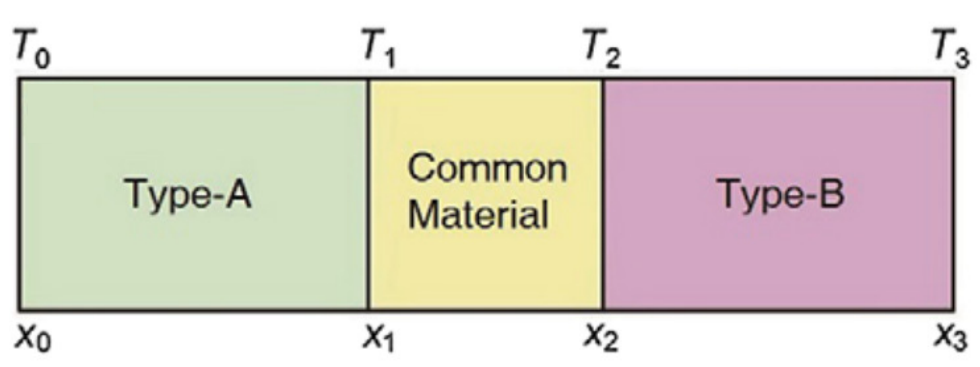}
\caption{\label{Fig. ZXC2}Schematic of an energy-free thermostat. Materials type-A and type-B are designed to regulate the temperature gradient from high to low. The region labeled ``common material'' is designated for temperature maintenance. Adapted from Ref. \cite{XCZhou-ShenPRL16}. With permission from the Author.}
 \end{figure}

Ref. \cite{XCZhou-ShenPRL16} introduced an energy-free thermostat for a conduction heat transfer system, as depicted in Fig. \ref{Fig. ZXC2}. This thermostat comprises three distinct materials: Type A, Type B, and a common material. The common material is designated as the functional region for temperature maintenance. Similar to designs found in other thermal metamaterials, the thermostat features a high-temperature boundary on the right ($T_\mathrm{H}$) and a low-temperature boundary ($T_\mathrm{L}$) on the left. Both the upper and lower boundaries are adiabatic. Impressively, this design aims to ensure that the temperature of the functional region remains stable under varying temperature gradients without any additional energy input.

The aforementioned functionality is achieved using the so-called ``temperature trapping theory." This theory originates from the one-dimensional Fourier's law, expressed as:
\begin{equation}
q=-\kappa\left(x,T\right)\frac{\mathrm{d}T}{\mathrm{d}x},
\end{equation}
where $q$ represents the heat flux, $\kappa$ denotes thermal conductivity, $T$ is the temperature, and $x$ indicates position. To maintain a consistent heat flux from the high-temperature boundary to the low-temperature boundary, represented as $q=C$, $\kappa (x,T)$ needs to be symmetric with $T=T_\mathrm{c}$. Furthermore, to ensure temperature uniformity in the central region, making $T_1\approx T_2 \approx T_\mathrm{c}$, there ought to be two pronounced temperature gradients in the regions $[x_0,x_1]$ and $[x_2,x_3]$, and a subdued gradient in the region $[x_1,x_2]$. As a result, the researchers determined the thermal conductivities of materials type-A and type-B as:
\begin{equation}
\label{Eq. ZXC2}
\left\{\begin{aligned}
&\kappa_\mathrm{A}=L(T_\mathrm{c}-T)=\delta+\frac{\varepsilon e^{T-T_\mathrm{c}}}{1+e^{T-T_\mathrm{c}}}\\
&\kappa_\mathrm{B}=L(T-T_\mathrm{c})=\delta+\frac{\varepsilon}{1+e^{T-T_\mathrm{c}}}
\end{aligned}
\right.,
\end{equation}
where $L$ is a logistic function dependent on $T$, $\delta$ is a small value, and $\varepsilon$ is a substantial value. By employing this strategy, the temperature is consistently maintained within the central region.

\begin{figure}[h!]
\centering
\includegraphics[width=6cm]  {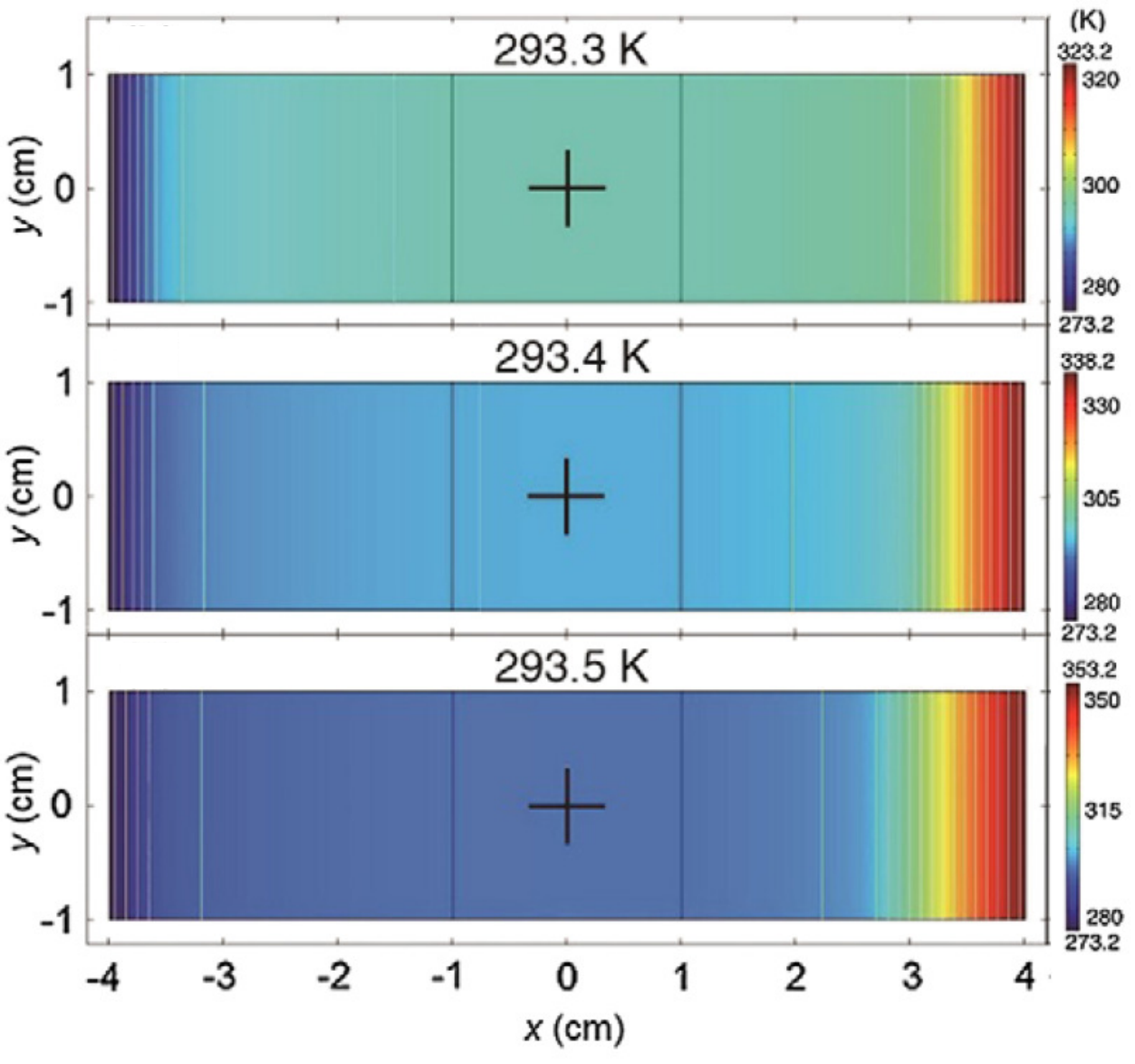}
\caption{\label{Fig. ZXC3}Simulation results for the energy-free thermostat. From the top subgraph to the bottom subgraph, the temperature at the low-temperature boundary stays consistent, while that of the high-temperature boundary incrementally rises. Adapted from Ref. \cite{XCZhou-ShenPRL16}. With permission from the Author.}
 \end{figure}

Figure \ref{Fig. ZXC3} presents a set of finite element simulations for the thermostat. Observably, with $T_\mathrm{L}$ held constant at $273.2\  \mathrm{K}$, and as $T_\mathrm{H}$ varies from $323.2\  \mathrm{K}$, $338.2\  \mathrm{K}$, to $353.2\  \mathrm{K}$, the temperature in the central region consistently remains within the range of $293.3-293.5$ K. This confirms the successful realization of the described temperature maintenance functionality.

\begin{figure}[h!]
\centering
\includegraphics[width=6cm]  {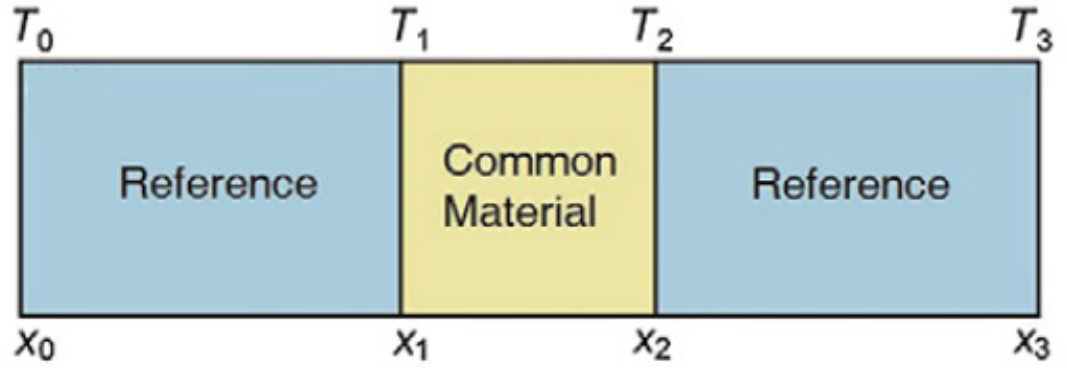}
\caption{\label{Fig. ZXC4} Schematic for a reference system. Adapted from Ref. \cite{XCZhou-ShenPRL16}. With permission from the Author.}
 \end{figure}

\begin{figure}[h!]
\centering
\includegraphics[width=6cm]  {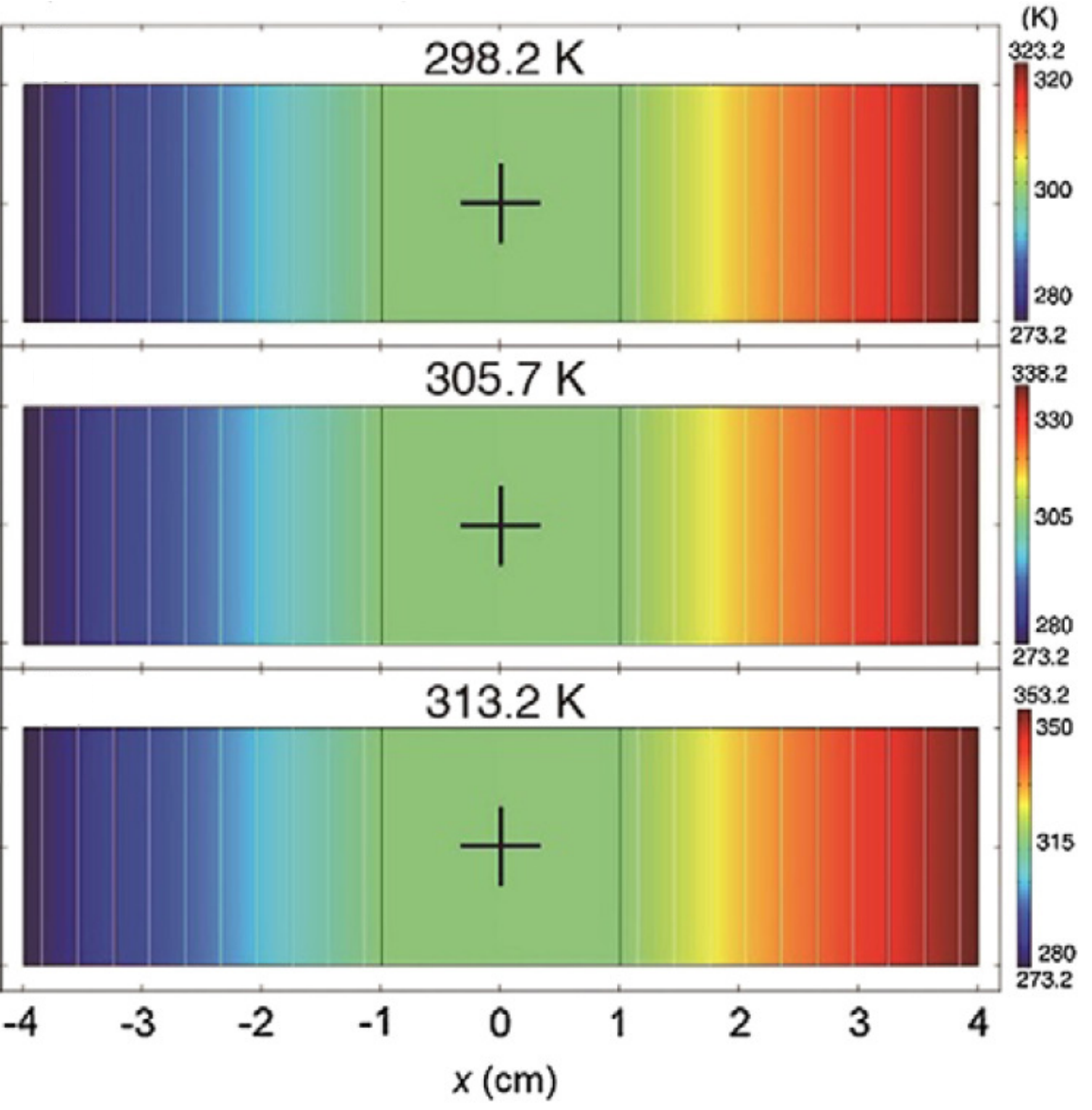}
\caption{\label{Fig. ZXC5} Simulation results for the reference system, serving as a control group for Fig. 2. Adapted from Ref. \cite{XCZhou-ShenPRL16}. With permission from the Author.}
 \end{figure}

For a comparative analysis, a reference system was established by substituting the type-A and type-B materials with materials having a constant thermal conductivity of $50\  \mathrm{W\  m^{-1}\  K^{-1}}$, as shown in Fig. \ref{Fig. ZXC4}. In the absence of the specialized designs of \(\kappa_\mathrm{A}\) and \(\kappa_\mathrm{B}\), the central region's temperature fails to remain stable under conditions identical to those of the energy-free thermostat, as depicted in Fig. \ref{Fig. ZXC5}. Notably, the temperature in the central region escalates in tandem with the rise in the ambient temperature gradient magnitude.

\begin{figure}[h!]
\centering
\includegraphics[width=12cm]  {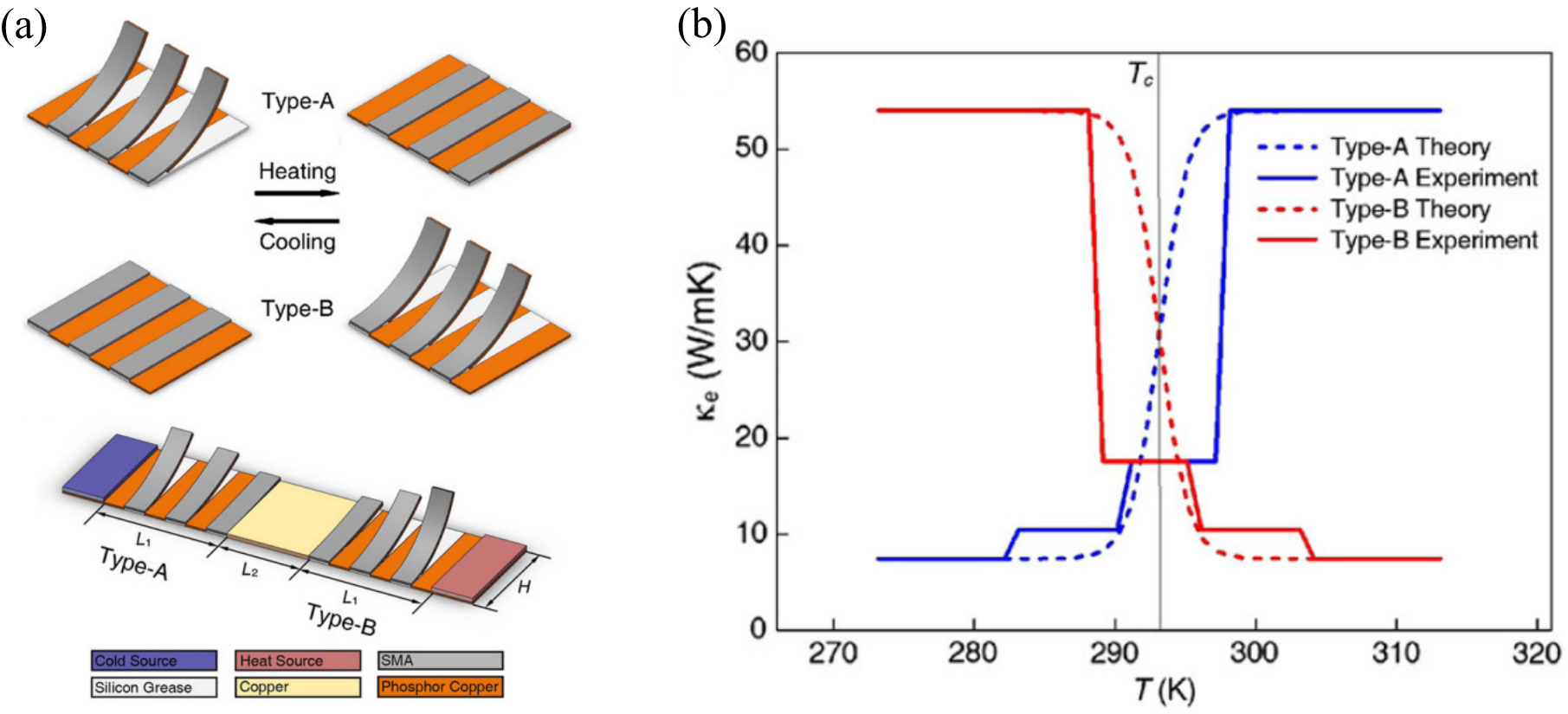}
\caption{\label{Fig. ZXC6} Schematic diagram of materials type-A and type-B. \textbf{a} Fabrication of the energy-free thermostat using shape memory alloy. \textbf{b} Comparison between theoretical and experimental results on the relationships of thermal conductivity and temperature for materials type-A and type-B. Adapted from Ref. \cite{XCZhou-ShenPRL16}. With permission from the Author.}
 \end{figure}

 While the ideal design of the energy-free thermostats demonstrates notable efficacy, the practical task of identifying suitable materials for type-A and type-B remains a significant challenge. To achieve the desired nonlinear thermal conductivity function, researchers employed a two-component system: a stationary component and a movable one. The movable component was realized using an SMA, while phosphor copper was chosen as the stationary component. This sophisticated structure is illustrated in Fig. \ref{Fig. ZXC6}a. When the temperature of the SMA surpasses its phase transition threshold, it experiences deformation. Capitalizing on this behavior facilitates the embodiment of the previously discussed function.
 
\begin{figure}[h!]
\centering
\includegraphics[width=10cm]  {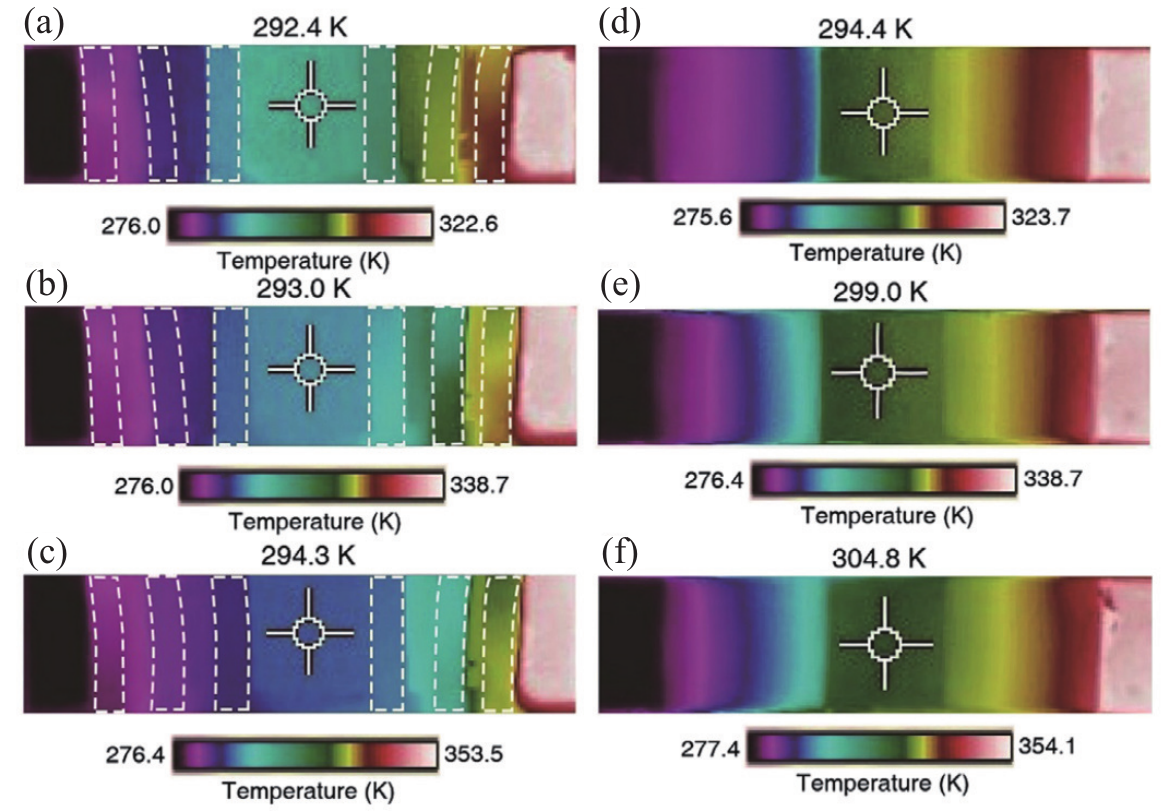}
\caption{\label{Fig. ZXC7} Experimental results showcasing the system's temperature distributions under various ambient temperature gradients. \textbf{a-c} Energy-free thermostat. \textbf{d-f} Control group. From the top subgraph to the bottom one, the temperature at the low-temperature boundary remains consistent, while the temperature at the high-temperature boundary incrementally rises. Adapted from Ref. \cite{XCZhou-ShenPRL16}. With permission from the Author.}
 \end{figure}

 A juxtaposition between the theoretical projections in Eq. (\ref{Eq. ZXC2}) and the empirical results of the composite structure, as shown in Fig. \ref{Fig. ZXC6}b, confirms that the design captures the targeted nonlinear properties. To validate the performance of the proposed energy-free thermostat, the researchers carried out experiments based on the configuration detailed in Fig. \ref{Fig. ZXC6}a, as shown in Fig. \ref{Fig. ZXC7}. The stable temperature observed in the central region, even amidst fluctuating external temperature gradients, hints at the feasibility of realizing the energy-free thermostat in real-world settings. Conversely, in scenarios where the SMA was excluded from the thermostat setup, the design's nonlinear thermal conductivity characteristics were compromised. As a result, the temperature in the central zone escalated from $294.4\ \mathrm{K}$ to $299.0\ \mathrm{K}$, and subsequently to $304.8\ \mathrm{K}$, in response to an increasing ambient temperature gradient.

\subsubsection{Comparison with existing technology}

The temperature trapping theory, combined with the design of energy-free thermostats, offers a revolutionary approach to efficient temperature maintenance. When juxtaposed with prevailing industrial solutions such as vapor compression refrigeration, thermal insulation, or phase-change temperature control, this strategy emerges as a quantitative, energy-saving alternative. It intricately regulates heat flow, obviating the need for energy input, and thus, perpetually sustains the desired temperature within the central region. This approach is especially promising for environments with fluctuating temperature gradients, be it the lunar surface or the contrasting sunlit and shaded facades of edifices \cite{XCZhou-HuangM2022}.

\begin{figure}[h!]
\centering
\includegraphics[width=8cm]  {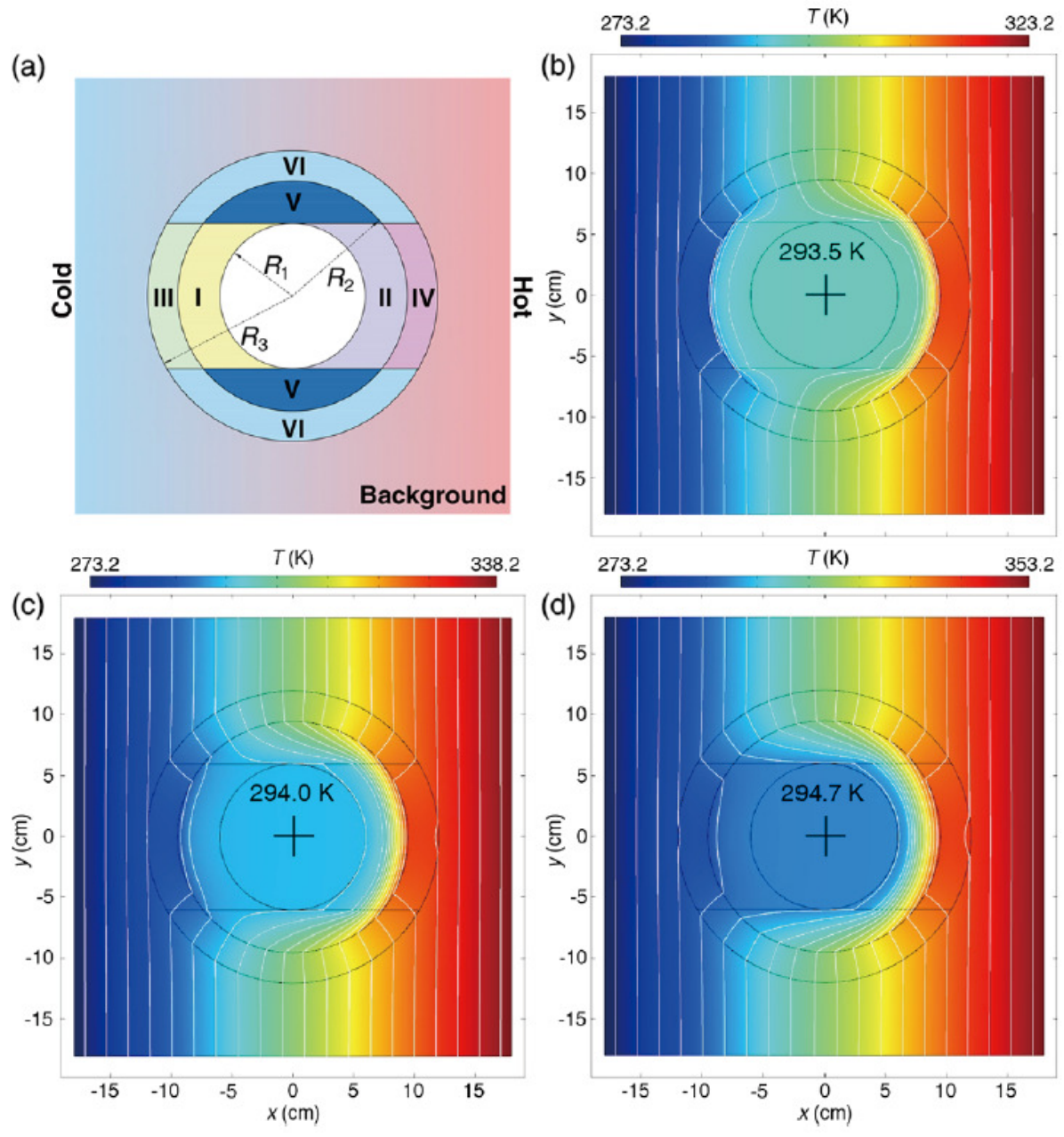}
\caption{\label{Fig. ZXC8} Improved thermal cloak with temperature maintenance capability. \textbf{a} Schematic representation. \textbf{b-d} Temperature distributions under various ambient temperature gradients. Adapted from Ref. \cite{XCZhou-ShenPRL16}. With permission from the Author.}
 \end{figure}

Furthermore, researchers harnessed this structure to enhance the inherent functionalities of thermal metamaterials, as depicted in Fig. \ref{Fig. ZXC8}. Defining the thermal conductivities of zones \uppercase\expandafter{\romannumeral1}-\uppercase\expandafter{\romannumeral6} as
\begin{equation}
\left\{\begin{aligned}
&\kappa_\mathrm{\uppercase\expandafter{\romannumeral1}}=\kappa_0-\frac{(\kappa_0-\kappa_\mathrm{i})}{1+e^{T-T_\mathrm{c}}},\\
&\kappa_\mathrm{\uppercase\expandafter{\romannumeral2}}=\kappa_0-\frac{(\kappa_0-\kappa_\mathrm{i})e^{T-T_\mathrm{c}}}{1+e^{T-T_\mathrm{c}}},\\
&\kappa_\mathrm{\uppercase\expandafter{\romannumeral3}}=\kappa_0+\frac{(\kappa_\mathrm{e}-\kappa_0)}{1+e^{T-T_\mathrm{c}}},\\
&\kappa_\mathrm{\uppercase\expandafter{\romannumeral4}}=\kappa_0+\frac{(\kappa_\mathrm{e}-\kappa_0)e^{T-T_\mathrm{c}}}{1+e^{T-T_\mathrm{c}}},\\
&\kappa_\mathrm{\uppercase\expandafter{\romannumeral5}}\rightarrow0,\\
&\kappa_\mathrm{\uppercase\expandafter{\romannumeral6}}=\kappa_0\frac{R_3^2+R_2^2}{R_3^2-R_2^2},
\end{aligned}
\right.
\end{equation}
it is discerned that the classical core-shell structure can now encapsulate the energy-free thermostat's capabilities. Fig. \ref{Fig. ZXC8}b-d elucidates the temperature maintenance attributes of a thermal cloak, for instance. As traditionally expected of thermal cloaks, the exterior temperature distribution around the shell remains unperturbed \cite{XCZhou-Tan16, XCZhou-FanAPL2008, XCZhou-NarayanaPRL2012, XCZhou-Maldovan2013, XCZhou-SchittnyPRL2013, XCZhou-MaPRL2014, XCZhou-XuPRL2014, XCZhou-HanPRL2014}. This innovative design preserves this characteristic, while simultaneously ensuring that, irrespective of shifting temperature gradients, the central region's temperature remains steadfast, thus augmenting the existing suite of thermal cloak functionalities \cite{XCZhou-YaoISci22,XCZhou-DaiPRAP22,XCZhou-WangPRAP21,XCZhou-WangATE21,XCZhou-WangEPL21,XCZhou-XuIJHMT21,XCZhou-XuCPL20,XCZhou-XuPRE20,XCZhou-YangJAP20,XCZhou-XuPRAP19,XCZhou-YangESEE19,XCZhou-YangJAP19,XCZhou-DaiJAP18,XCZhou-WangJAP18,XCZhou-DaiPRE18,XCZhou-Tan4,XCZhou-QiuIJHT14,XCZhou-QiuEPL13,XCZhou-GaoJAP2009,XCZhou-XuEPL201,XCZhou-LiJJAP10}.

\subsection{Negative-energy thermostat}\subsubsection{Principle and function}
\begin{figure}[h!]
\centering
\includegraphics[width=6cm]  {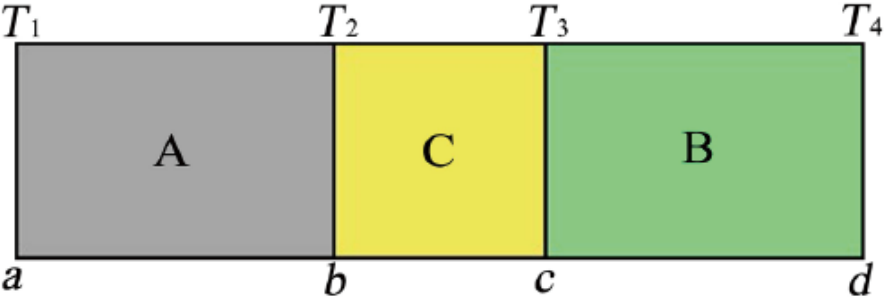}
\caption{\label{Fig. ZXC9} Schematic diagram of the negative-energy thermostat. The boundary conditions are analogous to those in Fig. 2. Adapted from Ref. \cite{XCZhou-WangPRA19}. With permission from the Author.}
 \end{figure}

Building upon the foundation of the energy-free thermostat, Ref. \cite{XCZhou-WangPRA19} introduced an innovative concept that harnesses thermoelectric effects to generate electricity, aptly termed the "negative-energy thermostat." The underlying schematic, when devoid of load, bears resemblance to that of the energy-free thermostat, as depicted in Fig. \ref{Fig. ZXC9}. While the temperature regulation mechanism echoes that of the energy-free thermostat, the ensuing discourse primarily elucidates the principles governing electricity control. Given that the bulk of the temperature gradient is confined to zones A and B, the thermoelectromotive force is described as follows:
\begin{equation}
\begin{aligned}
\Delta U_\mathrm{ad}&=\Delta U_\mathrm{ad}+\Delta U_\mathrm{cd}\\
&=S(T_2-T_1)+S(T_4-T_3)\\
&\approx S(T_4-T_1),
\end{aligned}
\end{equation}
where $S$ is the Seebeck coefficient.
 
In conjunction with the capabilities of the energy-free thermostat, it can be inferred that the potential for electricity output escalates as the ambient temperature gradient increases, all while the temperature of the central region remains stable.

\begin{figure}[h!]
\centering
\includegraphics[width=9cm]  {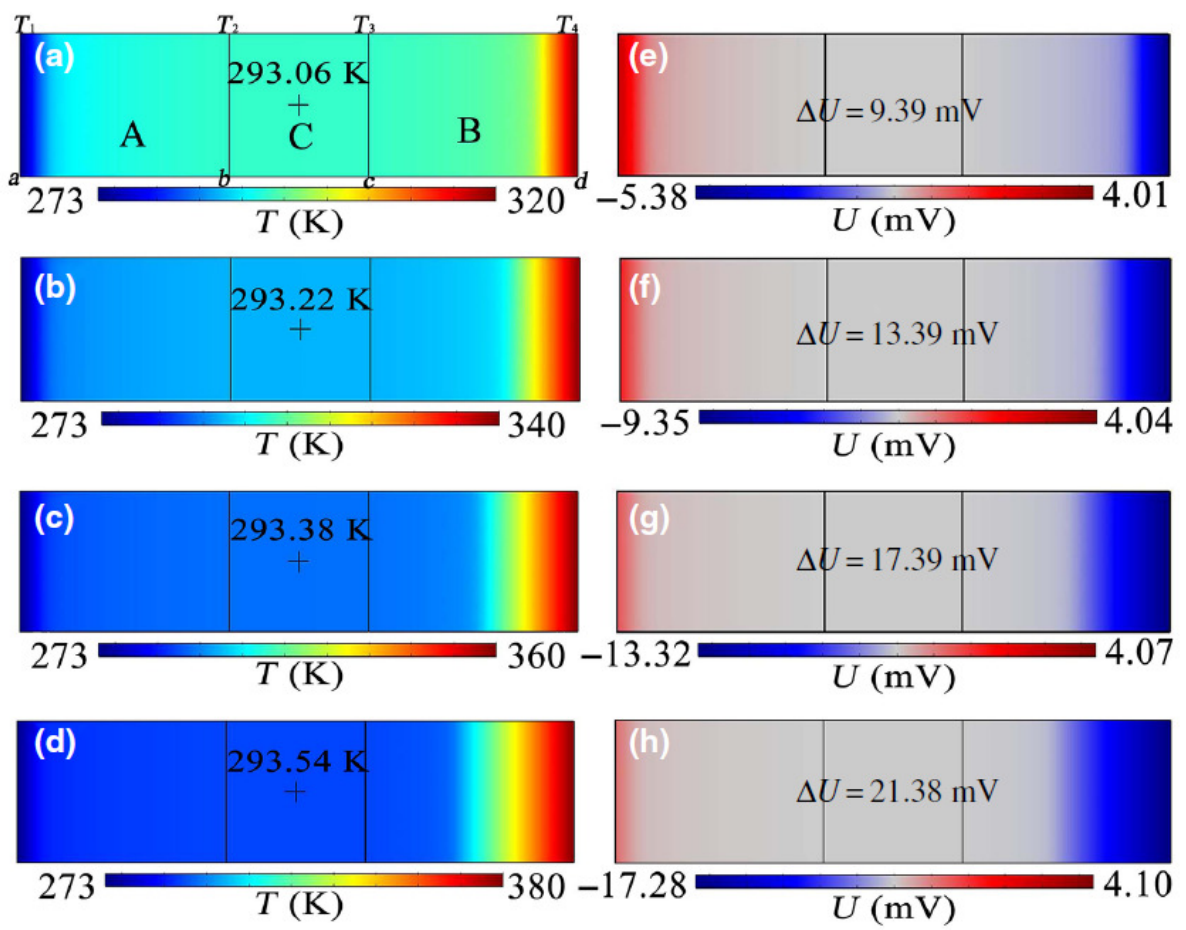}
\caption{\label{Fig. ZXC10} Finite element simulations of the negative-energy thermostat. \textbf{a-d} Temperature distributions and \textbf{e-h} electric potential distributions under various ambient temperature gradients. Adapted from Ref. \cite{XCZhou-WangPRA19}. With permission from the Author.}
 \end{figure}

Figure \ref{Fig. ZXC10} displays the finite element simulations, highlighting both the temperature maintenance performance and the electric potential of the negative-energy thermostat. Observations reveal that as the ambient temperature gradient intensifies, the central region's temperature is consistently maintained within the range of $293.06-293.54$ K. Simultaneously, $\Delta U$ experiences a surge, increasing from $9.39$ mV to $21.38$ mW. This underscores the promising application potential of the negative-energy thermostat.

In consideration of real-world applications, an external load is incorporated into the device. Given this modification, thermoelectrics come into play, necessitating a revision to the governing equation. The revised formulation was given by:

\begin{equation}
\begin{aligned}
&q(x)=-\kappa (T)\frac{\mathrm{d}T}{\mathrm{d}x}+\varPi j(x),\\
&j(x)=-\sigma \frac{\mathrm{d}\mu}{\mathrm{d}x}-\sigma S \frac{\mathrm{d}T}{\mathrm{d}x},
\end{aligned}
\end{equation}
where \(j(x)\) denotes the electric current density induced by the Seebeck effect, \(\mu\) represents the electrochemical potential at a given position, and \(\sigma\) signifies the electrical conductivity. \(\varPi\) is the Peltier coefficients and it is noted that \(\varPi=TS\).

\begin{figure}[h!]
\centering
\includegraphics[width=9cm]  {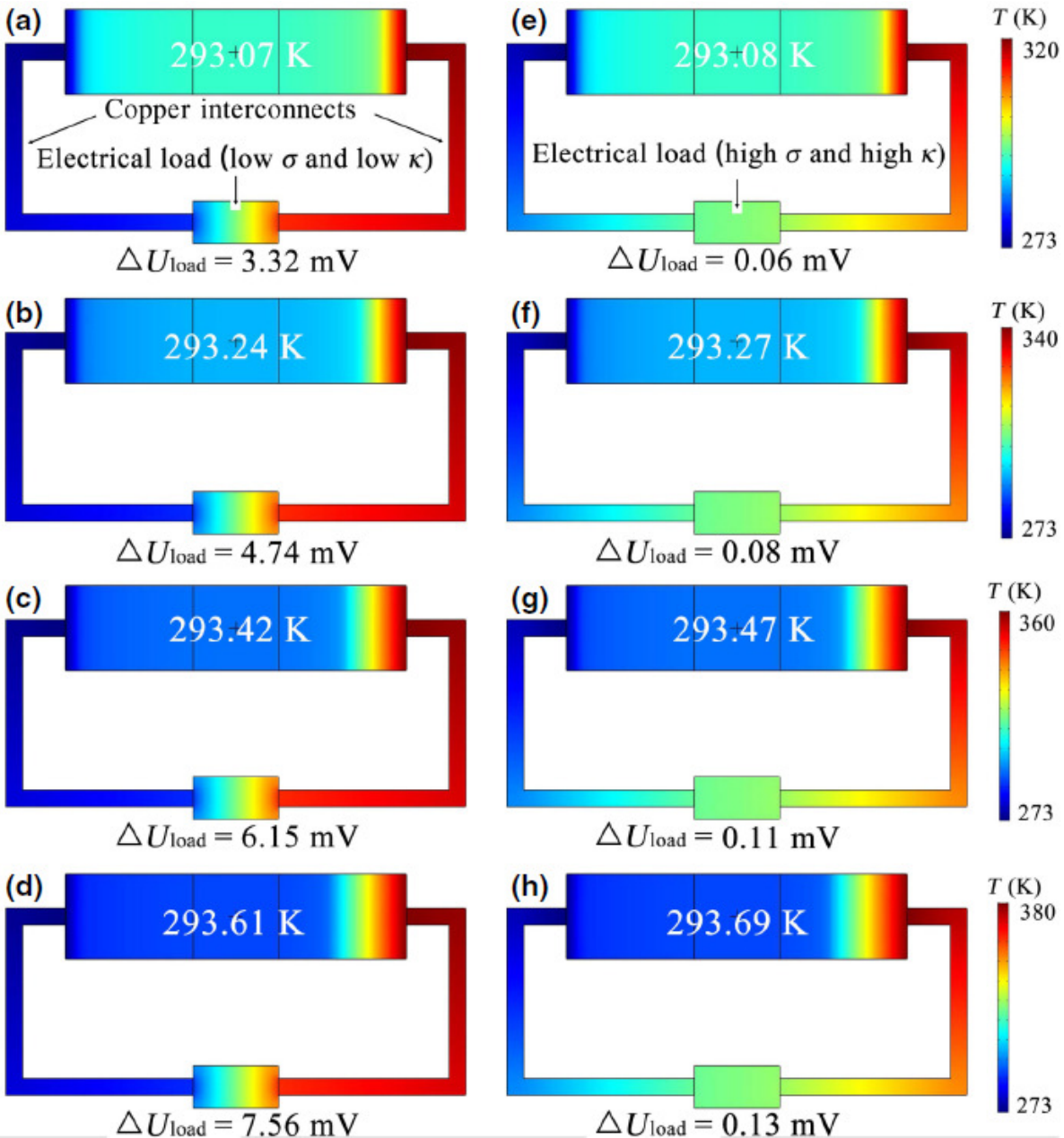}
\caption{\label{Fig. ZXC11} Finite element simulations showcasing the temperature distributions of the negative-energy thermostat with load under varying ambient temperature gradients. \textbf{a-d} Cases with low $\sigma$ and low $\kappa$. \textbf{e-h} Cases with high $\sigma$ and high $\kappa$.  Adapted from Ref. \cite{XCZhou-WangPRA19}. With permission from the Author.}
 \end{figure}

Figure \ref{Fig. ZXC11} displays the temperature distribution of the negative-energy thermostat, incorporating the external load, as portrayed in finite element simulations. As observed, an electrical load is connected to the original thermostat using copper. Two scenarios, concerning the thermal and electrical properties of the load, were examined. The first scenario focuses on low values of \(\sigma\) and \(\kappa\) (as shown in Fig. \ref{Fig. ZXC11}a-d), while the latter encompasses high \(\sigma\) and high \(\kappa\) values (depicted in Fig. \ref{Fig. ZXC11}e-h). The results show that both extreme conditions do not affect the central temperature of the thermostat.

\subsubsection{Comparison with existing technology}

\begin{figure}[h!]
\centering
\includegraphics[width=5cm]  {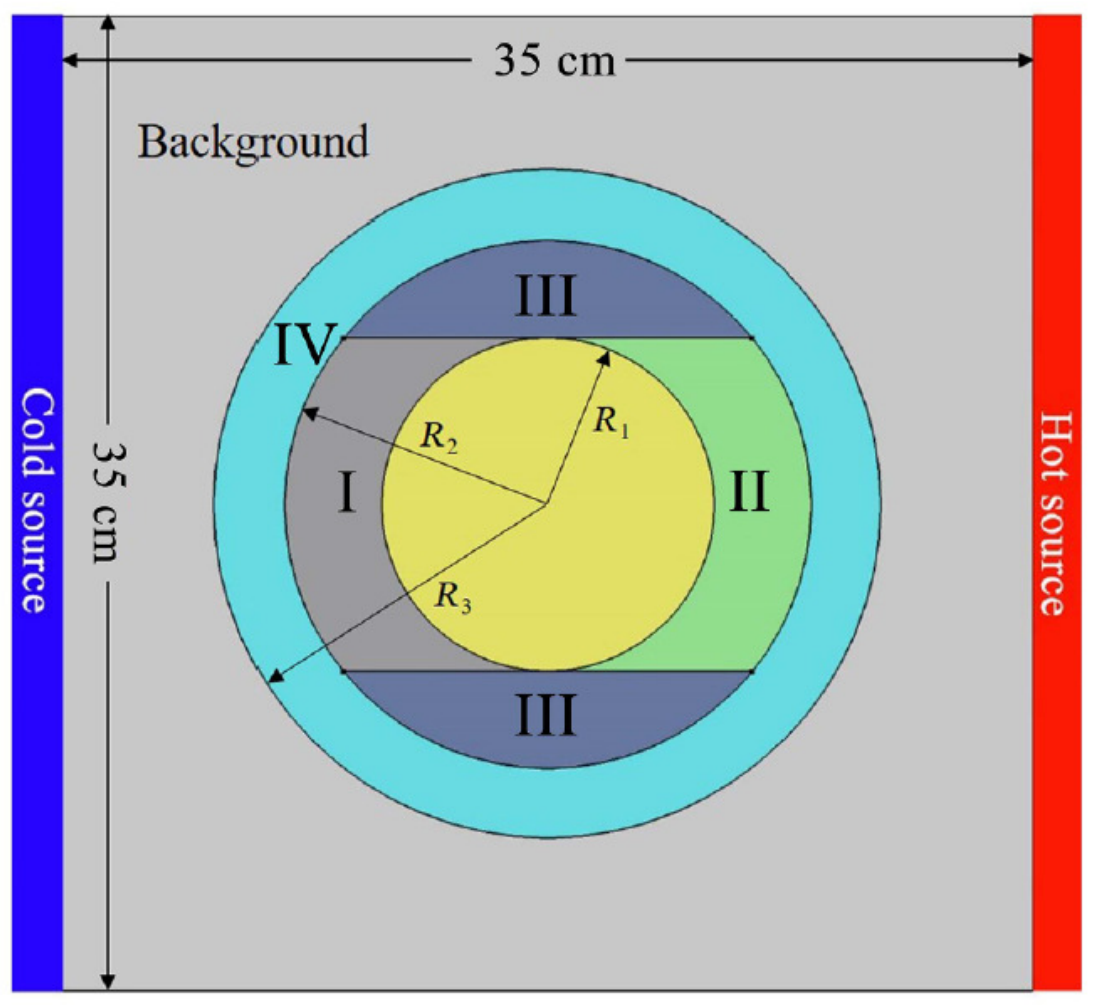}
\caption{\label{Fig. ZXC12} Schematic representation of a thermoelectric thermostat cloak. Adapted from Ref. \cite{XCZhou-WangPRA19}. With permission from the Author.}
 \end{figure}

\begin{figure}[htpb!]
\centering
\includegraphics[width=9cm]  {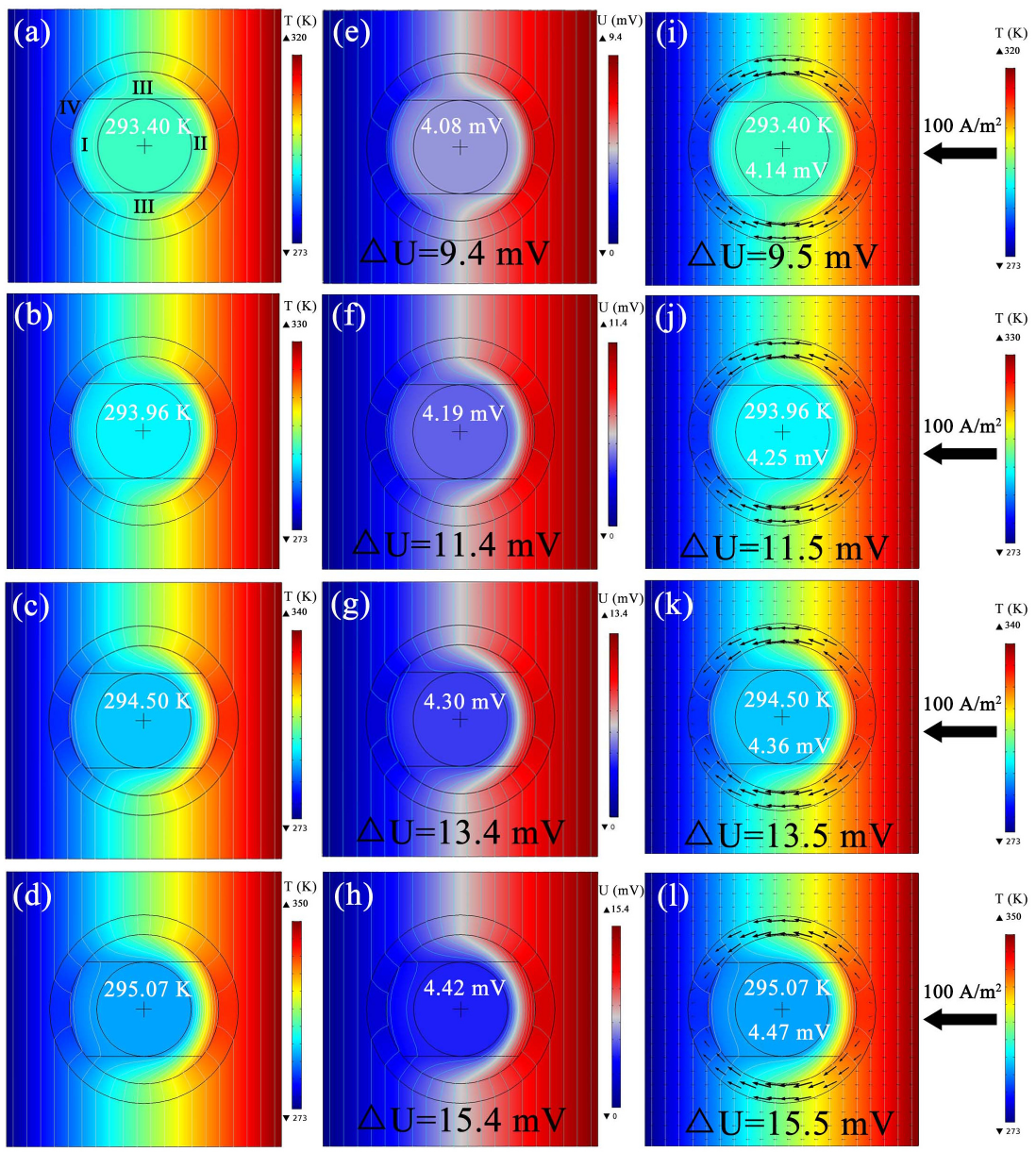}
\caption{\label{Fig. ZXC13} Finite element simulation of the thermoelectric thermostat cloak under various ambient temperature gradients. \textbf{a-d} Temperature distributions. \textbf{e-h} Thermoelectric potential distributions. \textbf{i-l} Temperature distribution featuring characteristic temperature and electric potential in the central region with input electric current. Adapted from Ref. \cite{XCZhou-WangPRA19}. With permission from the Author.}
 \end{figure}

Compared to the application of the energy-free thermostat in the thermal cloak, the negative-energy thermostat can be integrated into the thermoelectric cloak, as depicted in Fig. \ref{Fig. ZXC12}, endowed with the additional feature of electricity output. Drawing parallels with the functioning of the temperature-maintenance thermal cloak, researchers devised a thermoelectric cloak, illustrated in Fig. \ref{Fig. ZXC13}. The thermal and electrical conductivities for all regions were set as:
\begin{equation}
\left\{
\begin{aligned}
&\kappa_\mathrm{\uppercase\expandafter{\romannumeral1}}=\phi+\psi\frac{e^{T-T_\mathrm{c}}}{1+e^{T-T_\mathrm{c}}}\\
&\kappa_\mathrm{\uppercase\expandafter{\romannumeral2}}=\phi+\psi\frac{1}{1+e^{T-T_\mathrm{c}}}\\
&\kappa_\mathrm{\uppercase\expandafter{\romannumeral3}}\rightarrow 0\\
&\kappa_\mathrm{\uppercase\expandafter{\romannumeral4}}=\kappa_0\frac{R_3^2+R_2^2}{R_3^2-R_2^2}
\end{aligned}
\right.,
\end{equation}
and
\begin{equation}
\left\{
\begin{aligned}
&\sigma_\mathrm{\uppercase\expandafter{\romannumeral1}}=\mu+\nu\frac{e^{T-T_\mathrm{c}}}{1+e^{T-T_\mathrm{c}}}\\
&\sigma_\mathrm{\uppercase\expandafter{\romannumeral2}}=\mu+\nu\frac{1}{1+e^{T-T_\mathrm{c}}}\\
&\sigma_\mathrm{\uppercase\expandafter{\romannumeral3}}\rightarrow 0\\
&\sigma_\mathrm{\uppercase\expandafter{\romannumeral4}}=\sigma_0\frac{R_3^2+R_2^2}{R_3^2-R_2^2}
\end{aligned}
\right.,
\end{equation}
where $\phi$ and $\psi$ are parameters with the same units as $\kappa$, while $\mu$ and $\nu$ are parameters with the same units as $\sigma$ ($\phi \ll \psi$ and $\mu \ll \nu$).

\begin{figure}[htpb!]
\centering
\includegraphics[width=10cm]  {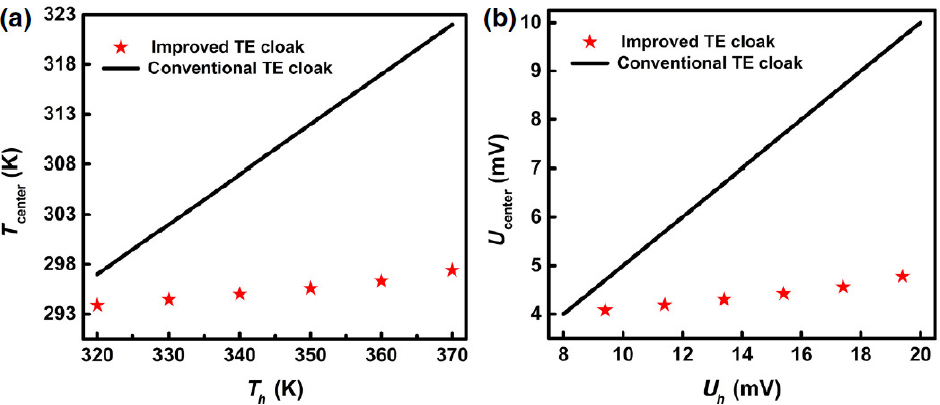}
\caption{\label{Fig. ZXC14}Comparison of the variations in \textbf{a} temperatures and \textbf{b} potentials for the thermoelectric thermostat cloak and the conventional thermoelectric cloak under varying ambient temperature and electric potential gradients. Adapted from Ref. \cite{XCZhou-WangPRA19}. With permission from the Author.}
\end{figure}

Initially, the thermoelectric cloak demonstrates an adeptness in maintaining the temperature within its central region. As evidenced by the temperature distribution in Fig. \ref{Fig. ZXC13}a-d, with the low-temperature boundary anchored at 273 K and the high-temperature boundary ranging from 320 K to 350 K, the temperature within the central region remains remarkably steady, fluctuating between 293.40 K and 395.07 K. Concurrently, the distribution of electrical potential follows a comparable trajectory, as depicted in \ref{Fig. ZXC13}e-h. Setting the baseline electrical potential boundary at 0 mV and escalating the high-end electrical potential boundary from 9.4 mV to 15.4 mV, the central region's electrical potential stabilizes within the 4.08-4.42 mV bracket. Intriguingly, this stability persists even in the presence of a current flowing from left to right, ensuring both the temperature and electrical potential remain constant (Fig. \ref{Fig. ZXC13}i-l). These findings underscore the enhanced functionalities of this modified thermoelectric cloak. Relative to the temperature-maintenance thermal cloak, it offers the additional advantage of electric potential regulation. Moreover, when set against a traditional thermoelectric cloak \cite{XCZhou-WangPRA19, XCZhou-StedmanTSR2017}, this advanced iteration notably curtails the temperature/electric potential deviations within the central region amid escalating ambient temperature/electric potential disparities, as illustrated in Fig. \ref{Fig. ZXC14}.


\subsection{Multi-temperature maintenance container}
\subsubsection{Principle and function}
The two previously mentioned types of thermostats employ SMA to regulate heat flow across a temperature control zone, ensuring its temperature remains consistent. Both in everyday scenarios and industrial applications, there is often a need for multi-temperature control. For instance, in indoor settings, different rooms might have specific temperature needs based on their functions. Similarly, for goods transportation, varying goods will have distinct temperature demands to preserve their quality. Ref. \cite{XCZhou-ZhouNC23} examined goods transportation and suggested an efficient multi-temperature control methodology.

\begin{figure}[h!]
\centering
\includegraphics[width=\textwidth]  {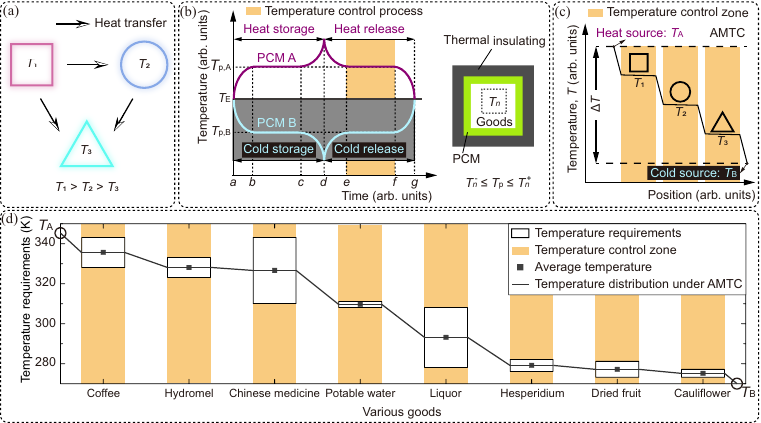}
\caption{\label{Fig. ZXC15} Multi-temperature control. \textbf{a} Heat transfer among various goods with different temperature requirements. \textbf{b} Principle of phase change temperature control and its utilization in cold chain logistics. \textbf{c} Schematic representation of adaptive multi-temperature control (AMTC). \textbf{d} Impact of AMTC in meeting multi-temperature requirements for eight goods. Adapted from Ref. \cite{XCZhou-ZhouNC23}. With permission from the Author.}
 \end{figure}

As depicted in Fig. \ref{Fig. ZXC15}a, when transported goods are at different temperatures (each with distinct temperature needs), heat transfer occurs amongst them. To ensure the safety of these goods, an efficient multi-temperature control mechanism is vital. An advanced temperature control method based on PCMs, commonly used in cold chain logistics, could offer guidance in this situation \cite{XCZhou-XXJFE21,XCZhou-ZYJES2020,XCZhou-LinNE2022}. This principle is illustrated in Fig. \ref{Fig. ZXC15}b. There are two types of PCMs utilized for temperature control. PCM A, termed the heat storage PCM, has a phase change temperature ($T_\mathrm{p,a}$) that's higher than the environmental temperature ($T_\mathrm{E}$). In contrast, PCM B, known as the cold storage PCM, has a phase change temperature ($T_\mathrm{p,b}$) below $T_\mathrm{E}$. The temperature-time curves can be divided into six stages based on their phase change processes. The segments $[a,b]\cup[c,d]$ signify the sensible heat (or cold) storage process. The segment [b,c] describes the latent heat (or cold) storage process. Meanwhile, $[d,e]\cup[f,g]$ indicates the sensible heat (or cold) release process, and $[e,f]$ represents the latent heat (or cold) release process. This method leverages the substantial latent heat storage capacity of the PCMs, controlling temperature through the release of latent heat (or cold). By packaging goods in PCMs and placing them in thermally insulating containers, the goods' temperature can be maintained over extended periods.

Yet, while utilizing PCMs to achieve multi-temperature control seems promising, the authors highlighted two primary challenges. First, it is crucial that the phase change temperature of the PCMs aligns with the goods' temperature needs, and they must also possess properties like thermal stability, chemical compatibility, cycle longevity, and cost-effectiveness \cite{XCZhou-XXJFE21,XCZhou-ZYJES2020,XCZhou-LinNE2022,XCZhou-ZhouNC23}. If suitable natural PCMs are not available, the creation of high-performance composite PCMs might be necessary, which might require a huge initial investment. Secondly, employing one PCM per zone and inserting thermal insulation materials does not entirely address the heat transfer issue between zones with varying temperature demands. Since the temperature range for goods transportation often surpasses that of cold chain logistics, the temperature disparity between zones can be significant, potentially requiring more energy to mitigate any multi-temperature control setbacks. Thus, the researchers aimed to devise a strategy to circumvent these challenges.

Their solution drew inspiration from terraced structures found in agricultural production, as portrayed in Fig. \ref{Fig. ZXC15}c. Recognizing that substances placed between hot and cold sources exhibit temperature drops, the team proposed creating multiple temperature platforms tailored for goods with diverse temperature requirements. This approach only necessitates the heat (or cold) source's temperature to be higher (or lower) than the goods with the most extreme temperature demands. Given the rapid advancements in phase change technology, selecting two high-performing PCMs to act as the primary hot and cold sources is feasible. Moreover, this strategy fully capitalizes on heat transfer between varying objects with diverse temperatures. Every object plays a part in facilitating multi-temperature control. Fig. \ref{Fig. ZXC15}d displays eight distinct goods, each with its temperature requirement. It demonstrates that using just one pair of hot and cold sources can naturally regulate the temperature of these eight items, underscoring the strategy's effectiveness. This strategy was termed as adaptive multi-temperature control (AMTC).
 
\begin{figure}[h!]
\centering
\includegraphics[width=9cm]  {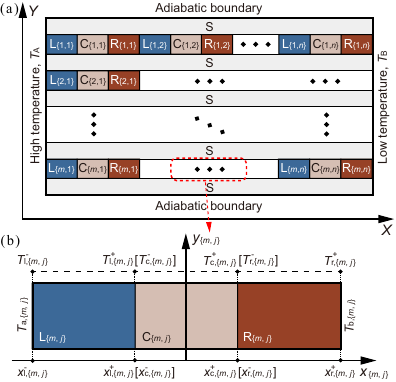}
\caption{\label{Fig. ZXC16} Multi-temperature control system. \textbf{a} Global system designed for multi-temperature control. \textbf{b} Local system located at the $m^\mathrm{th}$ row and $j^\mathrm{th}$ column. Adapted from Ref. \cite{XCZhou-ZhouNC23}. With permission from the Author.}
 \end{figure}

To achieve the previously discussed multi-temperature control with a terraced-shaped temperature distribution, the researchers developed a conduction heat transfer system, illustrated in Fig. \ref{Fig. ZXC16}. The left and right boundaries are set to high and low temperatures ($T_\mathrm{A}$ and $T_\mathrm{B}$), while the upper and lower boundaries are adiabatic. The system comprises four types of zones: $\mathrm{L}_{\{i,j\}}$,  $\mathrm{C}_{\{i,j\}}$, $\mathrm{R}_{\{i,j\}}$, and $\mathrm{R}$, where $i$ ($=1,2,3,\cdots,m$) symbolizes rows, and $j$ ($=1,2,3,\cdots,n$) represents columns. $\mathrm{C}_{\{i,j\}}$ designates the temperature control zone for storing goods. For effective multi-temperature control, researchers introduced thermal insulating materials into zone $\mathrm{S}$ and then moderated the temperature gradient in $\mathrm{L}_{\{i,j\}}$ and $\mathrm{R}_{\{i,j\}}$. The temperature in the control zone on column $m$ can be expressed as
\begin{equation}
\label{eq. 8}
\begin{aligned}
T_{\mathrm{c},{\{m,j\}}}&=T_\mathrm{A}-\frac{T_\mathrm{A}-T_\mathrm{B}}{\sum_{j^{\#}=1}^{n}{\sum_{\epsilon=\mathrm{l,c,r}}\frac{d_{\epsilon,\{m,j^{\#}\}}}{\kappa_{\epsilon,\{m,j^{\#}\}}}}}\Bigg(-\frac{d_{\mathrm{r},\{m,j\}}}{\kappa_{\mathrm{r},\{m,j\}}}\\
 &-\frac{d_{\mathrm{c},\{m,j\}}/2-x_{\{m,j\}}}{\kappa_{\mathrm{c},\{m,j\}}}+\sum_{j^*=1}^{j}\sum_{\epsilon=\mathrm{l,c,r}}\frac{d_{\epsilon,\{m,j^*\}}}{\kappa_{\epsilon,\{m,j^*\}}}\Bigg).
\end{aligned}
\end{equation}
In this equation, $\epsilon$ denotes zone $E$, where $\epsilon=\mathrm{l,c,r}$ corresponds to $E=\mathrm{L,C,R}$, respectively. For $j^{\#}=\{1,2,3\cdots n\}$ and $j^*=\{1,2,3\cdots j\}$, $\kappa_{\epsilon,\{m,j\}}$ and $d_{\epsilon,\{m,j\}}$ represent the thermal conductivity and the length traversed by heat flow in $E_{\{m,j\}}$, respectively. When $\kappa_{\mathrm{l},\{m,j^{\#}\}} \ll \kappa_{\mathrm{c},\{m,j\}}$ and $\kappa_{\mathrm{r}_,\{m,j^{\#}\}} \ll \kappa_{\mathrm{c},\{m,j\}}$, the temperature distribution in $\mathrm{C}_{\{i,j\}}$ is nearly uniform, allowing Eq. (\ref{eq. 8}) to be simplified as 
\begin{equation}
T_{\mathrm{c},{\{m,j\}}}\approx T_\mathrm{A}-\frac{T_\mathrm{A}-T_\mathrm{B}}{\sum_{j^{\#}=1}^{n}{\sum_{\epsilon=\mathrm{l,r}}\frac{d_{\epsilon,\{m,j^{\#}\}}}{\kappa_{\epsilon,\{m,j^{\#}\}}}}}\left(-\frac{d_{\mathrm{r},\{m,j\}}}{\kappa_{\mathrm{r},\{m,j\}}}+\sum_{j^*=1}^{j}\sum_{\epsilon=\mathrm{l,r}}\frac{d_{\epsilon,\{m,j^*\}}}{\kappa_{\epsilon,\{m,j^*\}}}\right).
\end{equation}
This suggests that the temperature in the control zone can be managed by adjusting the thermal or structural parameters of the system, specifically $\kappa_{\epsilon,\{m,j^{\#}\}}$ and $d_{\epsilon,\{m,j^{\#}\}}$.

\begin{figure}[h!]
\centering
\includegraphics[width=\textwidth]  {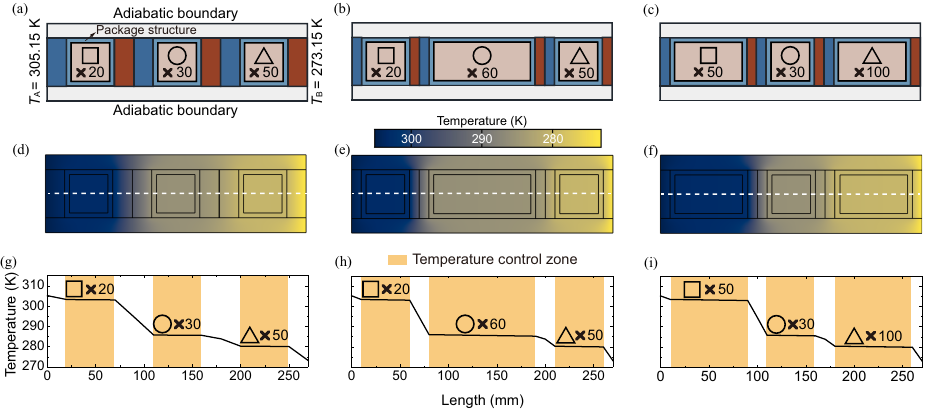}
\caption{\label{Fig. ZXC17} Multi-temperature control capability. Adapted from Ref. \cite{XCZhou-ZhouNC23}. \textbf{a-c} Various multi-temperature control systems designed for storing distinct goods in different quantities. \textbf{d-f} Finite element simulations illustrating the temperature distributions of the multi-temperature control system under steady-state conditions. \textbf{g-i} Terraced-shaped temperature profiles along the characteristic lines in \textbf{d-f}. Adapted from Ref. \cite{XCZhou-ZhouNC23}. With permission from the Author.}
 \end{figure}

This concept was validated through finite element simulations, as depicted in Fig. \ref{Fig. ZXC17}. Initially, the authors established a scenario wherein three types of goods required transportation, represented by square, circle, and triangle symbols. Their temperature requirements were defined as 303.15 K, 285.65 K, and 280.15 K, respectively. Fig. \ref{Fig. ZXC17}a-c illustrates the design of the multi-temperature control systems with varying structural parameters to accommodate these goods in different quantities. Adhering to the previously mentioned equation, the system's thermal parameters were computed, and the finite element simulations were executed, as presented in Fig. \ref{Fig. ZXC17}d-f. These figures demonstrate that the temperature distributions within the control zones were uniform. Additionally, to further evaluate the temperature control capability of the AMTC method, the authors studied the temperature distribution along the white characteristic lines. The findings, showcased in Fig. \ref{Fig. ZXC17}g-i, confirm that the terrace-shaped temperature distribution discussed earlier was effectively achieved by the AMTC, meeting the temperature needs of the goods impeccably.
 
\begin{figure}[h!]
\centering
\includegraphics[width=\textwidth]  {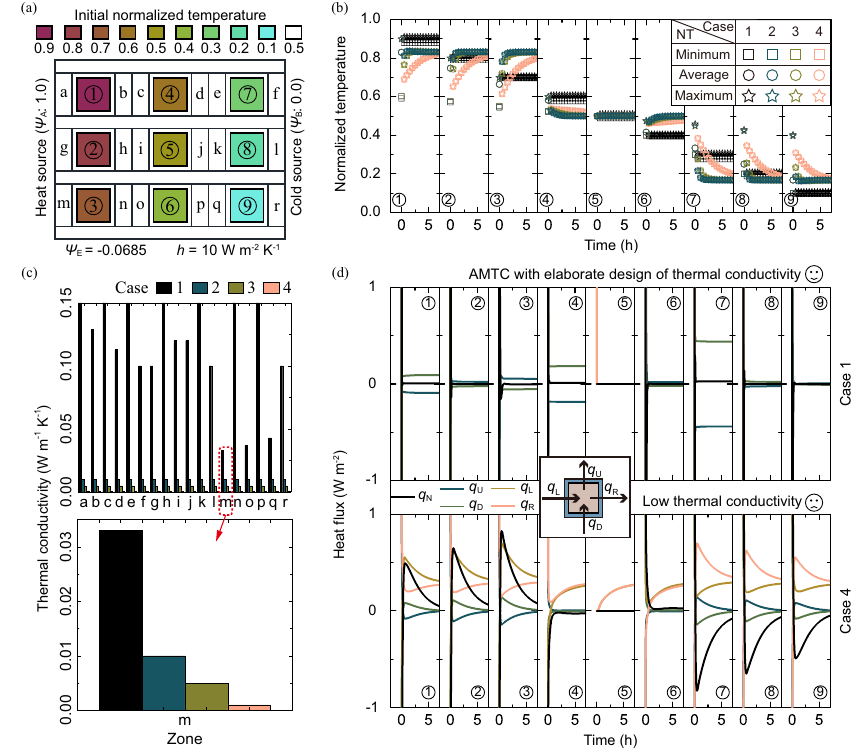}
\caption{\label{Fig. ZXC18} Multi-temperature maintenance. \textbf{a} Initial temperature distribution in the multi-temperature control system. \textbf{b} Temperature variations of objects within each temperature control zone for cases 1-4. \textbf{c} Thermal conductivities of the zones marked in \textbf{a}. \textbf{d} Heat flux analysis for each temperature control zone during case 1 and case 4.  Adapted from Ref. \cite{XCZhou-ZhouNC23}. With permission from the Author.}
 \end{figure}

Using the AMTC approach, the capability for multi-temperature maintenance was evaluated under transient heat transfer conditions, as depicted in Fig. \ref{Fig. ZXC18}. Figure \ref{Fig. ZXC18}a presents a scenario for transporting a variety of goods with distinct temperature requirements. For generality, the researchers employed normalized temperatures for this investigation. The temperatures of the heat and cold sources were defined as 1.0 and 0.0, respectively, while the initial temperatures of the goods in zones \textcircled{\scriptsize1}, \textcircled{\scriptsize2}, $\cdots$, \textcircled{\scriptsize9} were designated as 0.9, 0.8, $\cdots$, 0.1, respectively. The thermal conductivities of the marked zones a, b, $\cdots$, r were determined using the AMTC method. As illustrated in case 1 of Fig. \ref{Fig. ZXC18}b, after a brief initial disturbance to equalize temperature differences, the temperatures of all goods were effectively maintained by the multi-temperature control system. For comparison, three scenarios with diminished thermal conductivities were considered: $0.01\  \mathrm{W\  m^{-1}\  K^{-1}}$ for case 2, $0.005\  \mathrm{W\  m^{-1}\  K^{-1}}$ for case 3, and $0.001\  \mathrm{W\  m^{-1}\  K^{-1}}$ for case 4 (Fig. \ref{Fig. ZXC18}c). Observations indicate that while decreasing thermal conductivities appears beneficial for enhancing multi-temperature maintenance performance, even when conductivities were minimized to an extremely low value (as in case 4), the desired multi-temperature maintenance characteristics were not achieved, as presented in Fig. \ref{Fig. ZXC18}b. To delve deeper, the researchers conducted a heat flux analysis for each temperature control zone, showcased in Fig. \ref{Fig. ZXC18}d. Cases 1 and 4 were selected for comparison. The heat flux traversing the temperature control zone from the left, right, top, and bottom were labeled as $q_\mathrm{L}$, $q_\mathrm{R}$, $q_\mathrm{U}$, and $q_\mathrm{D}$, respectively. The net heat flux is expressed as $q_\mathrm{N}=q_\mathrm{L}-q_\mathrm{R}+q_\mathrm{D}-q_\mathrm{U}$, representing the intensity of heat exchange both inside and outside the temperature control zone. It is evident that in case 1, employing the AMTC strategy, the net heat flux approaches zero after a brief disturbance, which is consistent with the temperature variation behavior of the goods. However, in case 4, regardless of the extent to which thermal conductivities were reduced, the erratic heat flux across each temperature control zone compromised the multi-temperature maintenance performance.

\begin{figure}[h!]
\centering
\includegraphics[width=\textwidth]  {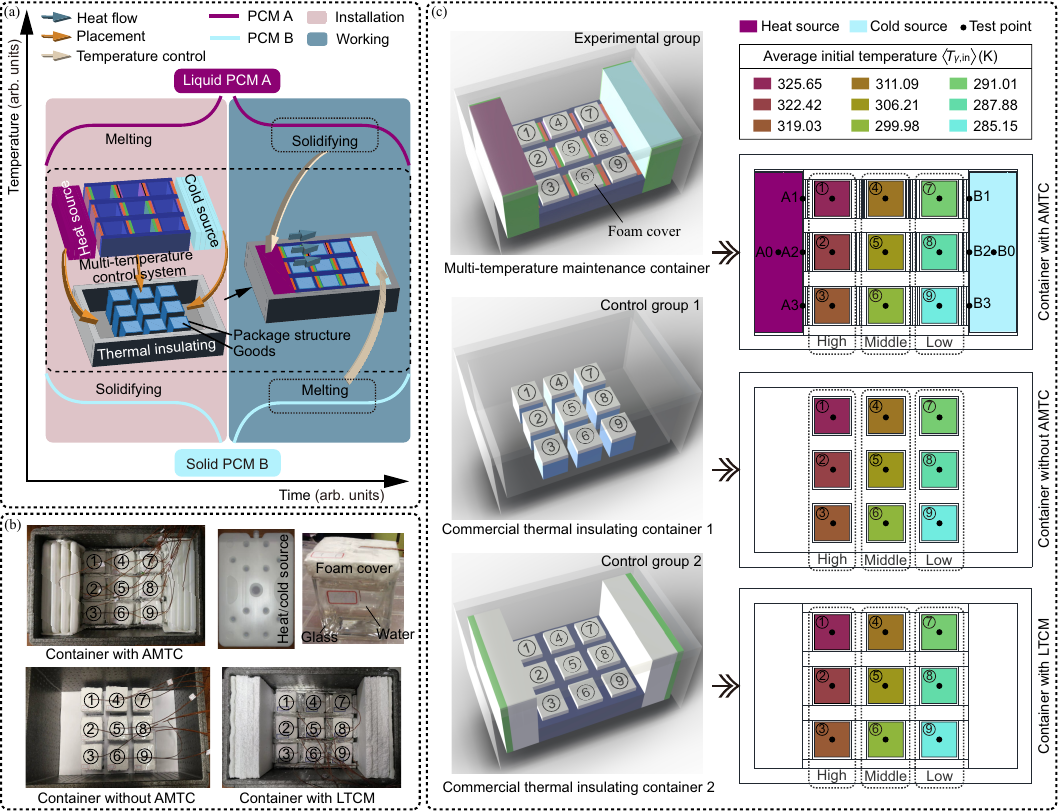}
\caption{\label{Fig. ZXC19} Multi-temperature maintenance container. \textbf{a} Schematic illustration of the operation mechanism. \textbf{b,c} Experimental setup and schematic representation for the container with AMTC (multi-temperature maintenance container), without AMTC (commercial thermal insulating container 1), and with LMTC (commercial thermal insulating container 2). AMTC stands for adaptive multi-temperature control, while LMTC represents low thermal conductivity materials. Adapted from Ref. \cite{XCZhou-ZhouNC23}. With permission from the Author.}
 \end{figure}

Building on the findings, the authors ventured to design a multi-temperature maintenance container suited for real-world goods transportation. The conceptual design is presented in Fig. \ref{Fig. ZXC19}a. This apparatus consists of four primary components: a pair of portable hot and cold sources, a multi-temperature control system, a standard insulation box, and nine compartments tailored to house items at diverse temperatures. The heat source is fashioned from stearic acid (PCM A), which has a phase-change temperature around 340.15 K, whereas the cold source is derived from distilled water (PCM B) with a phase-change temperature close to 273.15 K. Given that the thermal characteristics of the multi-temperature control system are derived from computational methods, it is conceivable that no naturally occurring materials might fulfill these specifications. To address this, the authors employed the effective thermal resistance approach, constructing the system by amalgamating various commercially available materials. For operation, the stearic acid needs to be thoroughly melted to serve as the heat reservoir, and the distilled water must be fully frozen to act as the cold reservoir. Positioned in their designated spots, these elements offer consistent high and low-temperature boundaries, facilitating multi-temperature regulation throughout the transportation process.

The tangible engineering prototype for the multi-temperature maintenance container is displayed in Fig. \ref{Fig. ZXC19}b and is conveniently termed the ``container with AMTC." To highlight the efficiency of the AMTC, the authors also crafted two control variants. One is a container without the AMTC feature (a conventional thermal insulating container, termed control group 1), while the other employs a container using materials with low thermal conductivity as a substitute for AMTC (designated as a commercial thermal insulating container, or control group 2). Figure \ref{Fig. ZXC19}c illustrates a three-dimensional representation of the aforementioned prototypes accompanied by corresponding sectional diagrams. Each distinct color in the temperature control zones denotes the initial temperature of the stored goods. For the sake of generality, water was utilized to emulate the goods, given its prototypical thermophysical properties and ease of initial temperature setup. The color bar provides a representation of the average initial temperatures.

\begin{figure}[h!]
\centering
\includegraphics[width=14cm]  {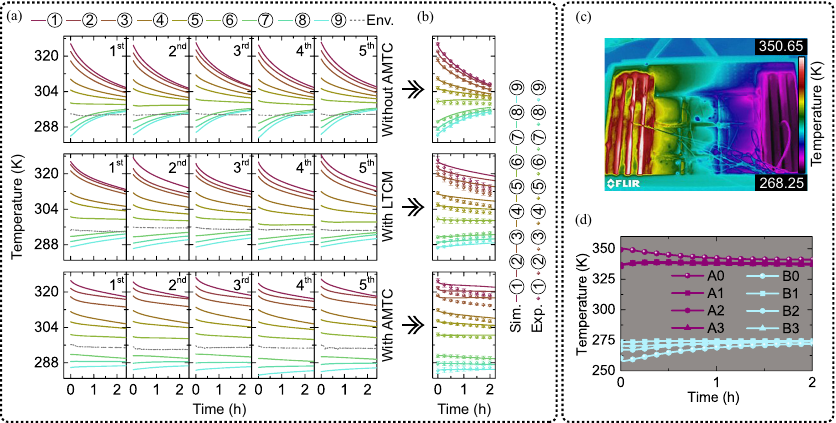}
\caption{\label{Fig. ZXC20} Multi-temperature maintenance performance. \textbf{a} Temperature-time curves for the simulacrum in each temperature control zone across five independent tests. \textbf{b} Comparison between experimental and simulation results for the temperature-time curves of each simulacrum. \textbf{c} Infrared image showing the temperature distribution inside the container after opening the lid. \textbf{d} Temperature-time curves for each test point corresponding to the mobile heat and cold sources. Adapted from Ref. \cite{XCZhou-ZhouNC23}. With permission from the Author.}
\end{figure}

The efficacy of multi-temperature maintenance provided by the aforementioned three devices is illustrated by the temperature-time curves of the simulacra, as presented in Fig. \ref{Fig. ZXC20}. Fig. \ref{Fig. ZXC20}a showcases five independent tests, whereas Fig. \ref{Fig. ZXC20}b contrasts simulation results with the average values derived from experimental outcomes. When juxtaposing the trajectories of the temperature-time curves from the container without AMTC, with LTCM, and with AMTC, a discernible observation is that the temperature fluctuation rates of the simulacra housed in the container with AMTC are the most subdued. This underscores the beneficial impact of the AMTC method in enhancing multi-temperature maintenance performance. Additionally, a separate evaluation was conducted for the mobile heat and cold sources, depicted in Fig. \ref{Fig. ZXC20}c-d. The steady temperature they furnish aligns with the requirements for high-/low- temperature boundaries as illustrated in Fig. \ref{Fig. ZXC16}.

\begin{figure}[h!]
\centering
\includegraphics[width=14cm]  {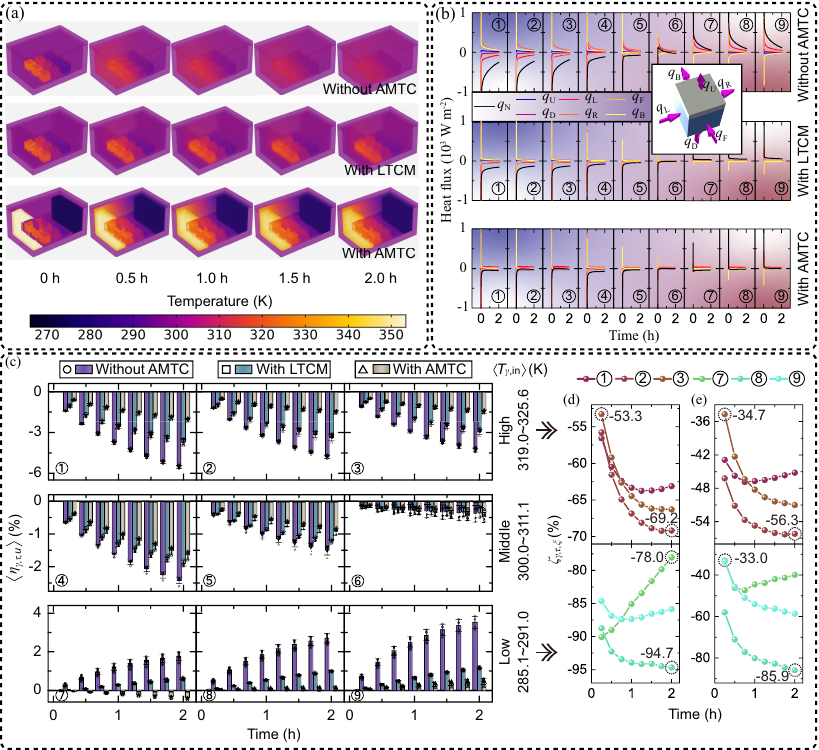}
\caption{\label{Fig. ZXC21} Effect of AMTC on multi-temperature maintenance performance. \textbf{a} Temperature distributions inside the container without AMTC, with LTCM, and with AMTC, at specific times. \textbf{b} Heat flux analysis for each temperature control zone. \textbf{c} Temperature variation rates for the simulacrum in each temperature control zone. \textbf{d,e} Changes in temperature variation rates of the simulacrum in the high- (\textbf{d}) and low-temperature (\textbf{e}) regions.  Adapted from Ref. \cite{XCZhou-ZhouNC23}. With permission from the Author.}
 \end{figure}

Following an in-depth analysis of the multi-temperature maintenance mechanisms, finite element simulations were utilized to visually elucidate temperature gradients within the storage compartments across specific time intervals, as depicted in Fig. \ref{Fig. ZXC21}a. The comprehensive heat flux assessment for each designated temperature zone highlighted that in the absence of AMTC, erratic temperature fluxes between simulacra of disparate temperatures impeded the proficient maintenance of multiple temperature zones, especially in regions with pronounced temperature extremes, as evidenced in Fig. \ref{Fig. ZXC21}b. Subsequently, a quantitative metric, $\left\langle\eta_{\gamma,\tau,u}\right\rangle$, was formulated to gauge the average temperature variation rate of the simulacrum within the container, as showcased in Fig. \ref{Fig. ZXC21}c.

Insights revealed that over two hours, the magnitude of temperature deviation, $\left|\left\langle\eta_{\gamma,\tau,u}\right\rangle\right|$, for simulacra contained in the AMTC-equipped storage ranged narrowly between 0.14\% and 2.05\%. A comparative scrutiny among containers devoid of AMTC, those outfitted with LTCM, and those incorporating AMTC, accentuated a correlation: amplified temperature fluctuations inversely influenced the efficiency of multi-temperature maintenance, underscoring the pivotal role of AMTC. Delving deeper, in high- and low-temperature regions with extreme temperature variations, the deviation, $\left|\left\langle\eta_{\gamma,\tau,u}\right\rangle\right|$, for containers employing AMTC was curtailed by 53.3\%-69.2\% and 78.0\%-94.7\% respectively in contrast to their non-AMTC equivalents, and by 34.7\%-56.3\% and 33.0\%-85.9\% respectively when contrasted against LTCM-integrated containers, as illuminated in Fig. \ref{Fig. ZXC21}d,e. Such pronounced performance enhancement attributed to AMTC underscores its indispensable utility, particularly for the reliable transport of sensitive goods.

\subsubsection{Comparison with existing technology}
Integrating the principles of thermal metamaterials' heat flow modulation with phase change mechanisms, the development of multi-temperature maintenance container offers innovative solutions to pivotal societal issues in the realm of dependable goods transportation. While most contemporary temperature regulation strategies prioritize meticulous heat flow control within a singular domain in the above energy-free thermostat and negative-energy thermostat, the presented methodology champions the establishment of multiple zones, each having tunable temperatures. This not only augments its pragmatic aspect but also accentuates its portability and storage-centric attributes.

When benchmarked against prevalent temperature maintenance techniques that employ phase change modalities to engineer multi-temperature insulating chambers for cold chain management, often employing composite PCMs of optimal phase change temperatures, the devised strategy in this study curtails the necessity for extensive hot and cold source deployment. Furthermore, it amplifies the scope of temperature modulation domains. Such an enhancement translates to a streamlined operational protocol and a more economical upfront investment. In essence, this research harmoniously marries core principles of physics with actionable insights from engineering, paving the way for a simplified yet more versatile temperature modulation framework.
\section{Prospects for convection heat transfer system}
The preceding sections presented three specific cases demonstrating temperature maintenance with diverse functionalities within a conduction heat transfer system. Following the developmental trajectory in thermal metamaterials, the logical next step is to explore temperature maintenance in a convection heat transfer system. Given the ubiquity of convection heat transfer in various scenarios, advancing in this direction will foster broader applications in the field of temperature maintenance.

From the discussed operations, two pivotal factors emerge in the fabrication of actual devices. First is the SMA characterized by variable structural parameters, and the second is the phase transition of liquid-solid PCMs. Both elements, individually or in tandem with thermal metamaterials, are instrumental in achieving meticulous heat flow regulation. The pertinent question then arises: can this precise control be replicated in a convection heat transfer setting?

\begin{figure}[h!]
\centering
\includegraphics[width=11cm]  {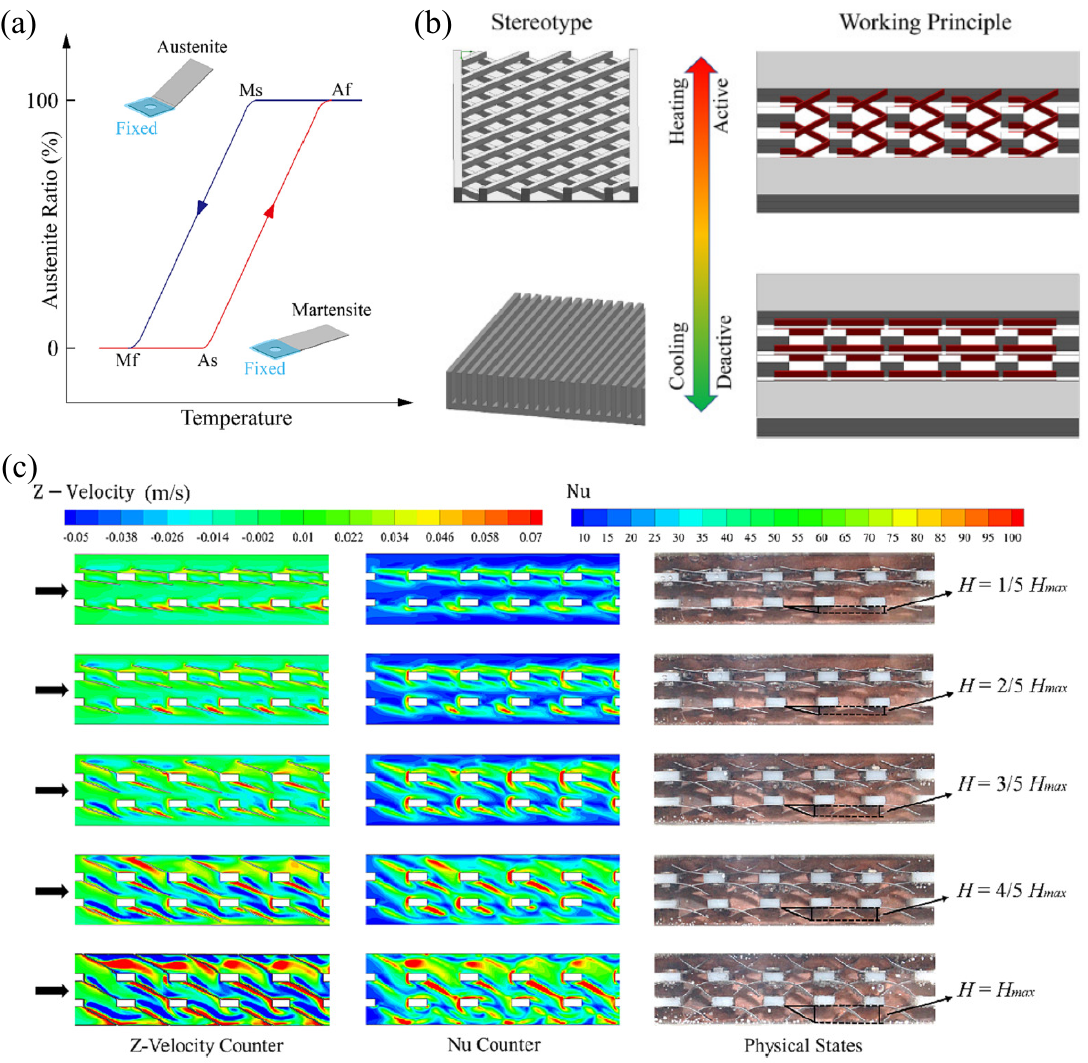}
\caption{\label{Fig. ZXC22} Smart heat exchanger based on SMA. \textbf{a} Ideal process in martensite-austenite transitions of NiTi. \textbf{b} Working principle of the smart heat exchanger. \textbf{c} Simulation results under various physical states. Adapted from Ref. \cite{XCZhou-WangBATE23}. \textcircled{c} 2022 Elsevier.}
 \end{figure}

Citing SMA as a reference, Ref. \cite{XCZhou-WangBATE23} unveiled a cutting-edge heat exchanger underpinned by SMA. SMA, conceptualized as the linchpin, automatically modulates the convective heat transfer performance. The mechanism driving the SMA employs NiTi as the foundational material. Its capability to oscillate between austenite and martensite phases ensures that the SMA can toggle between active and inactive states, as delineated in Fig. \ref{Fig. ZXC22}a. Crafting the heat exchanger, as illustrated in Fig. \ref{Fig. ZXC22}b, the objective was to leverage the SMA's transformative capacity to alter the flow channel's architecture. In cooler temperatures, the SMA remains dormant, but as temperatures soar, it becomes active. Analyzing the flow dynamics, it is evident that the activated state induces fluid turbulence, bolstering heat transfer. Fig. \ref{Fig. ZXC22}c showcases the simulated outcomes for this intelligent exchanger, revealing a direct correlation: as the SMA warping intensifies, the local Nusselt (Nu) number climbs, signaling enhanced convective heat transfer.

Conversely, employing PCMs primarily aims to stabilize the temperature within a designated region. Numerous studies have showcased the adeptness of PCMs in modulating temperature amidst convective heat transfer—examples span applications in aircraft, vehicles, and even in systems that store heat energy to subsequently convert it into electricity \cite{XCZhou-ElefsiniotisAJEM2013, XCZhou-RenXMATE2020}.

While both these technologies exhibit the prowess to qualitatively manage temperature under convection heat transfer, thermal metamaterials bring to the table a quantitative toolkit rooted in analytical theory. Integrating thermal metamaterials into this domain could be a potential move, amplifying the precision and scope of temperature control.

\section{Summary}
This review delves into three innovative temperature maintenance devices rooted in the advancements of conduction heat transfer, all harnessing the potential of thermal metamaterials: the energy-free thermostat, the negative-energy thermostat, and the multi-temperature maintenance container. The energy-free thermostat stands out with its capability to stabilize the temperature in the central region under fluctuating temperature gradients without necessitating any additional energy. The negative-energy thermostat, on the other hand, can uphold both temperature and electrical potential in the central area of thermoelectric cloak amidst varying temperature and electrical potential gradients, all while producing electrical energy. The third, the multi-temperature maintenance container, offers the versatility of concurrently maintaining different objects, each with unique temperature requirements, a feature especially pertinent for goods transportation applications. The design methodologies underpinning these devices may well pave the way for future innovations in passive temperature control technology. Moreover, the principles of heat flow regulation, informed by thermal metamaterials, are extended to encompass situations involving convection heat transfer. Building upon existing pioneering passive temperature control techniques in pertinent research domains, this review concludes by offering a glimpse into the promising future of temperature maintenance, further empowered by thermal metamaterials. 

In recent research, thermal metamaterials have been extended to chemical systems \cite{XCZhou-ZhangATS22}, plasma systems \cite{XCZhou-ZhangCPL22}, and more, collectively termed as diffusion metamaterials \cite{XCZhou-YangFBRMP23,XCZhou-ZhangNRP23}. Drawing inspiration from a variety of systems in physics, chemistry, and engineering, such as astrophysics \cite{ZhangPRD22}, fluid dynamics \cite{XCZhou-LiPF22, LiuIJMPB11}, phase-transition systems \cite{GaoNM21,Tan3,Tan7,Tan12,Tan19,QiuCTP15,ZhangCPB10,HanPLA2009,GuCTP2009}, water science \cite{MengCPB18,Tan8,Tan14,Tan17,QiuJP.Phys.Chem.B15,QiuTEPJ-AP14,Meng2013pre,MengJPCB11,TianEPL07}, complex systems \cite{Tan2,Tan5,liang2013pre,SongARCS12,ZhaoPNAS11,ChenJPA07,QiuCPB14,YePA081}, electromagnetism \cite{Tan18,QiuCTP14,Wang2013ACM,WangOL10,FanCTP10,TianJAP2009,ZhangCPL2009,2008,ZhangAPL08,TianPRE07,FanAPL06,TianCPL06,WangJPCB06,HuangPRE05a,HuangPLA05, HuangPRE04d,HuangJPCB04,KoJPCM04,HuangCTP03,HuangCTP02,HuangPRE02,HuangJPCM02}, nonlinear science \cite{XuPLA06,HuangJAP06, HuangAPL05a,HuangJPCB05,HuangEL04,HuangPRE04e,HuangPRE01, PanPB01,HuangSSC00}, colloidal systems \cite{Li2013EPJP,Fan2013cpb, chen2013prl,Li2012sm,GaoPP10,BaoJPCM10,GaoPRL10,WuEPJAP09,XiaoJPCB2008,JianJRCB08,XuJMR08,HuangJRCC08,ZhuJAP08, GaoJPCC07,FangCPL07,FanJPCB06,CaoJPCB06,ShenCPL06,WangCPL06,TianPRE06,HuangPRE05b,LiuCTP05,HuangPRE04a,HuangPRE04b, HuangPLA04,HuangJPCM04,HuangPRE04c,HuangJCP04, HuangCP04, HuangPRE04f,HuangCPL04, LiuPLA04,HuangPRE03-1, HuangJAP03, GaoPRE03, HuangPRE03-2, HuangPLA02, HuangJMMM05,TanJPCB20091,DongJAP041}, nanoscience \cite {Meng2013mp,Wang2012cpb,WangJPCB11,ZhaoJAP2009,GaoAPL2008}, optics \cite{Fan2013fop,FanJPCC2009,JianJPCC2009,WangOL2009,WangOL2008, FanJAP2008,HuangNY07,WangAPL07,HuangAPL05b,XiaoPRB05,HuangCTP01-2}, acoustics \cite{huang2013cp,ZhaoFOP12,FanJPDAP11,SuFOP11, LiuCTP10, LiuEPJAP2009}, soft matter \cite{TanSM10,TanJPCB2009}, graded materials \cite{HuangOL05, HuangJAP05, HuangAPL04,KoEPJE04,HuangPRE04g, DongJAP04, GaoEPJB03}, and others novel mechanisms/functions \cite{HuangFP17, XinPA17, Tan10, XueCPL06, DongEPJB05, ZhangCPL23,LiuLPO2013,JinPRe2023,TanHPRE2023}, the principle of temperature maintenance could be further refined and made more advantageous for practical applications.

\end{document}